\documentclass[review]{elsarticle}%

\usepackage{amsmath}
\usepackage{amsfonts}
\usepackage{amssymb}
\usepackage{amsthm}
\usepackage{bm}
\usepackage{indentfirst}
\usepackage{graphicx}
\usepackage{subfigure}
\usepackage{array}
\usepackage{fullpage}

\DeclareMathAlphabet{\mathsfsl}{OT1}{cmss}{m}{sl}

\newcommand{\PreserveBackslash}[1]{\let\temp=\\#1\let\\=\temp}
\newcolumntype{C}[1]{>{\PreserveBackslash\centering}p{#1}}
\newcolumntype{R}[1]{>{\PreserveBackslash\raggedleft}p{#1}}
\newcolumntype{L}[1]{>{\PreserveBackslash\raggedright}p{#1}}

\numberwithin{equation}{section}

\theoremstyle{definition}

\usepackage{anyfontsize}
\usepackage{enumerate}
\usepackage{esint}
\usepackage{etoolbox}
\usepackage{isomath}
\usepackage{mathrsfs}
\usepackage{mathtools}
\usepackage{multicol}
\usepackage{pgfplots}
\usepackage{scalerel}
\usepackage{stackengine}
\usepackage{tabulary}
\usepackage{tikz}

\makeatletter
\newcommand*\bdot{\mathpalette\bdot@{.65}}
\newcommand*\bdot@[2]{\mathbin{\vcenter{\hbox{\scalebox{#2}{$\m@th#1\bullet$}}}}}
\makeatother

\makeatletter
\newcommand*\bddot{\mathpalette\bddot@{.65}}
\newcommand*\bddot@[2]{\mathbin{\vcenter{\hbox{\scalebox{#2}
    {$\m@th#1\smash{{}_{\bullet}^{\bullet}}$}}}}}
\makeatother

\newcommand{\circled}[2][]{%
  \tikz[baseline=(char.base)]{%
    \node[shape = circle, draw, inner sep = .5pt]
    (char) {\phantom{\ifblank{#1}{#2}{#1}}};%
    \node at (char.center) {\makebox[0pt][c]{#2}};}}
\robustify{\circled}

\stackMath
\newcommand\reallywidecheck[1]{%
\savestack{\tmpbox}{\stretchto{%
  \scaleto{%
    \scalerel*[\widthof{\ensuremath{#1}}]{\kern-.6pt\bigwedge\kern-.6pt}%
    {\rule[-\textheight/2]{1ex}{\textheight}}
  }{\textheight}%
}{0.5ex}}%
\stackon[1pt]{#1}{\scalebox{-1}{\tmpbox}}%
}
\usepackage[utf8]{inputenc}
\usepackage{bm}
\usepackage{mathrsfs}
\usepackage{amsmath}
\usepackage{amsfonts}
\usepackage{amsthm}
\usepackage[ruled]{algorithm} 
\usepackage[noend]{algpseudocode} 
\usepackage{color}
\usepackage{setspace}
\usepackage{float}
\usepackage{import}
\usepackage{url}
\usepackage{multicol}
\usepackage{multirow}
\usepackage{subfigure}
\biboptions{sort&compress}
\usepackage[top=1in,bottom=1in,left=1in,right=1in]{geometry}

\usepackage{tikz}

\def\halfcheckmark{\tikz\draw[scale=0.4,fill=black](0,.35) -- (.25,0) -- (1,.7) -- (.25,.15) -- cycle (0.75,0.2) -- (0.77,0.2)  -- (0.6,0.7) -- cycle;}

\def\cmark{\tikz\draw[scale=0.4,fill=black](0,.35) -- (.25,0) -- (1,.7) -- (.25,.15) -- cycle (0.75,0.2);}

\newcommand{\real}{\mathbb{R}}

\newcommand{\mcD}{\mathcal{D}}

\newcommand{\mcG}{\mathcal{G}}

\newcommand{\mcJ}{\mathcal{J}}

\usepackage{xcolor}

\def\omg{{\Omega}}

\def \bb{\bm{b}}

\def \fb{\bm{f}}

\def \ub{\bm{u}}

\def \wb{\bm{w}}
\def \vb{\bm{v}}
\def \xb{\bm{x}}

\def \hb{\bm{h}}

\def \Kb{\bm{K}}

\def \qb{\bm{q}}
\def \nb{\bm{n}}
\def \yb{\bm{y}}
\def \cb{\bm{c}}
\def \Cb{\bm{C}}
\def \eb{\bm{e}}

\def \xib{{\boldsymbol\xi}}
\def \etab{{\boldsymbol\eta}}

\def \mbF{\mathbb{F}}
\def \mbU{\mathbb{U}}

\newcommand{\ubA}{{\underline{\bm A}}}

\newcommand{\calB}{{\mathcal{B}}}

\newcommand{\ue}{{\underline e}}
\newcommand{\ubM}{{\underline{\bm M}}}
\newcommand{\ubT}{{\underline{\bm T}}}
\newcommand{\bv}{{\bm v}}

\newcommand{\ut}{{\underline t}}
\newcommand{\ubY}{{\underline{\bm Y}}}
\newcommand{\Rb}{{\bm R}}
\newcommand{\uomega}{{\underline\omega}}
\newcommand{\uone}{{\underline 1}}
\newcommand{\ux}{{\underline x}}

\newcommand{\vertii}[1]{{\left\vert\left\vert #1
    \right\vert\right\vert}}    
    
\newcommand{\verti}[1]{{\left\vert #1
    \right\vert}}

\usepackage{lineno}
\usepackage[hidelinks]{hyperref}

\begin{document}
\begin{frontmatter}

\title{Peridynamic Neural Operators: A Data-Driven Nonlocal Constitutive Model for Complex Material Responses}

\address[yy]{Department of Mathematics, Lehigh University, Bethlehem, PA, USA}
\address[ss]{Center for Computing Research, Sandia National Laboratories, Albuquerque, NM, USA}
\address[nl]{Global Engineering and
Materials, Inc., Princeton, NJ, USA}
\address[zz]{Department of Mathematical Sciences, Worcester Polytechnic Institute, Worcester, MA, USA}

\author[yy]{Siavash Jafarzadeh}
\author[ss]{Stewart Silling}
\author[nl]{Ning Liu}
\author[zz]{Zhongqiang Zhang}
\author[yy]{Yue Yu\corref{cor1}}\ead{yuy214@lehigh.edu}

\cortext[cor1]{Corresponding author}


\begin{abstract}
Neural operators, which can act as implicit solution operators of hidden governing equations, have recently become popular tools for learning the responses of complex real-world physical systems. Nevertheless, most neural operator applications have thus far been data-driven and neglect the intrinsic preservation of fundamental physical laws in data. In this work, we introduce a novel integral neural operator architecture called the Peridynamic Neural Operator (PNO) that learns a nonlocal constitutive law from data. This neural operator provides a forward model in the form of state-based peridynamics, with objectivity and momentum balance laws automatically guaranteed. As applications, we demonstrate the expressivity and efficacy of our model in learning complex material behaviors from both synthetic and experimental data sets. We show that, owing to its ability to capture complex responses, our learned neural operator achieves improved accuracy and efficiency compared to baseline models that use predefined constitutive laws. Moreover, by preserving the essential physical laws within the neural network architecture, the PNO is robust in treating noisy data. The method shows generalizability to different domain configurations, external loadings, and discretizations.
\end{abstract}


\end{frontmatter}

\tableofcontents

\section{Introduction}
The detection and prediction of complex material responses from experimental spatial measurements are often needed in applications of interest to the broad scientific and engineering community \cite{zohdi2002toughening,wriggers1998computational,prudencio2013dynamic,su2006guided,AFOSR2014,talreja2015modeling,soric2018multiscale,pijaudier2013damage,mourlas2019accurate,markou2021new}. In disciplines ranging from structural health prognosis to non-destructive evaluation, material characteristics need to be accurately captured to guarantee reliable and trustworthy predictions. For many decades, constitutive models within PDE-based continuum mechanics have been commonly employed to characterize material response. Typically in such characterization, a strain energy density function is predefined with a specific functional form. Then, the material parameters are calibrated through an inverse method or analytical stress--strain fitting to test data. The outcome is a predefined model with fitted material parameters that can be employed as a homogenized surrogate for a real specimen in applications, such as the prediction of material deformation under a broad range of loading scenarios.

The parameter-fitting PDE approach faces three challenges that can affect the reliability of the resulting surrogates.
First, the descriptive power of predefined constitutive models is often restricted to certain deformation modes and ranges of strain. This tends to limit the predictivity and generalizability of the model \cite{lee2014inverse,he2021manifold,lee2017vivo}. 
Second, the required functional form of the expert-constructed model is not always available. 
For example, in spite of many years of research, there is not a definitive material model for many biotissues such as skin \cite{limbert2019skin,tacc2023benchmarking}. 
Third, if the fitted material model is a homogenized surrogate for a heterogeneous material, it inevitably approximates the original problem only in an averaged sense \cite{milton2022theory}. 
As a result, it may ``wash out'' the length scale in the original material microstructure, causing information to be lost that can be important in multiscale material modeling problems.

To address the first of these challenges and obtain more flexible predictive models, machine learning (ML) based approaches have emerged as potential solutions. 
Many of the successful prior attempts on data-driven material modeling can be classified as one of the following two types: 
(1) approaches that learn a constitutive law \cite{ghaboussi1998new,tacc2023benchmarking,zhang2020machine,vlassis2020geometric,liu2020generic,fuhg2022learning}, and 
(2) model-free learning approaches \cite{li2022fourier,goswami2022physics,yin2022simulating,yin2022interfacing,you2022physics,you2022learning,liu2023ino,liu2023domain,liu2023clawno,goswami2022pino}. 
The first approach aims to obtain the strain energy potentials, often in terms of the sums of convex non-decreasing functions of invariants and linear combinations of them, by fitting stress-strain curves. 
Additional physical requirements, such as objectivity and polyconvexity, are imposed in the design. 
Recent examples include input convex neural networks (ICNN) \cite{klein2022polyconvex}, constitutive artificial neural networks (CANN) \cite{linka2023new}, and neural ordinary differential equations (NODE) \cite{tac2022data}. 
However, since these approaches learn from stress-strain curves, they face challenges in capturing material heterogeneity.
It can also be difficult for these approaches to handle more general types of training data beyond stress-strain curves, such as displacement measurements from mesoscale simulations \cite{you2022data} and full-field Digital Image Correlation (DIC) \cite{chu1985applications} measurements.

The second ML approach, model-free learning, offers more flexibility, because the material response can be modeled as a mapping between loading conditions and the resulting displacement and damage fields \cite{goswami2022pino,goswami2022physics,you2022physics,you2022learning,liu2023ino,liu2023domain}. 
In this approach, the constitutive law can be seen as a hidden PDE, and the target mapping serves as the solution operator for this PDE. The goal is then to train a neural network based on spatial measurements as a surrogate for the implicit solution operator. 
Different neural network architectures have been proposed within this general approach, including the deep neural networks \cite{yang2022deep,de2018multidimensional,yang2020learning}, graph neural networks \cite{guo2020semi,xue2023learning,hestroffer2023graph,thomas2023materials,gladstone2023gnn}, and convolutional neural networks \cite{jin2022dynamic,kaviani2023high,holm2020overview}, and others. However, the model-free approach models generally do not guarantee fundamental physical laws. As a result, their performances highly rely on the quantity and coverage of available data. Moreover, the learnt network is not generalizable to different grids and domain shapes. That means, modeling the material response for a new instance of new domain shape with a different discretization often requires a new neural network model and correspondingly training from scratch.


Recently, the nonlocal neural operator was proposed as a way to learn a surrogate mapping between function spaces \cite{li2020neural,lanthaler2023nonlocal}.  
With this type of operator, the increment between layers is modeled as a nonlocal (integral) operator to capture the long-range dependencies in the feature space. 
The nonlocal neural operator can therefore be seen as a stack of discretized integral equations.
When the number of discretization nodes changes, the discretization still corresponds to the same continuous integral equation, and therefore the optimal network parameters can be reused. 
As a result, the operators feature resolution independence.
They also serve  as universal approximators to any function-to-function mapping operator \cite{lanthaler2023nonlocal}. 
These advantages make them good candidates for discovering models for complex material response directly from spatial measurements. In \cite{you2022physics}, neural operators were employed to construct a surrogate mapping from each loading function to its corresponding material response function directly from digital image correlation (DIC) measurements. However, although they provide generalizability to different resolutions, the vanilla neural operators are not generalizable to different domain shapes \cite{liu2023domain,li2023geometry}. On the other hand, same as other model-free neural networks, nonlocal neural operators rely on data to obtain the intrinsic physical laws, and hence they are prone to performance deterioration due to overfitting and measurement noises, especially in the small and noisy data regime \cite{lu2021comprehensive,zhang2023metano}.

Beyond their advantages in neural network designs, {\emph{nonlocal}} operators are also currently receiving much attention as homogenized surrogate models that  efficiently and accurately capture small-scale effects \cite{du2020multiscale}. 
In comparison with classical PDE models, these operators can better incorporate the coupling between constituent materials in a composite microstructure, taking into account the length scales in the microstructure (see Section~\ref{sec:exp} later in this paper for examples of this). 
The presence of such length scales in a heterogeneous material leads to technologically important observable effects such as wave dispersion in real materials.
Nonlocality also emerges from certain approaches to homogenization. 
Since nonlocal models usually include a finite cut-off distance for interactions (the {\emph{horizon}}), 
an intrinsic length scale such as the length of polymer chains can be introduced into the model in a natural way \cite{maranganti2007length}.
Nonlocal models have therefore been proposed as homogenized surrogates for molecular dynamics, providing efficient upscaled models that bridge nanoscale material structures and homogenized continuum responses \cite{beran1970mean,cherednichenko2006non,karal1964elastic,rahali2015homogenization,neuman2009perspective}.

Nevertheless, in previous work, the form of the nonlocal constitutive model has been adopted on a case-by-case basis, informed by either experimental evidence \cite{diehl2019review,benson2000application} or analytical derivations that help match the desired physical properties \cite{jiang2023spatiotemporally,deshmukh2022multiband,difonzo2023physics}. In prior work \cite{You2021,you2021data,lu2023nonparametric,fan2023bayesian,zhang2022metanor}, our team has developed the nonlocal operator regression approach, which learns the nonlocal kernel of an integral equation from data. 
This approach has been effective in providing homogenized coarse grained models for graphene sheets from molecular dynamics simulations \cite{you2022data,you2023towards}.
It has also succussfully reproduced features of stress wave propagation in composites and metamaterials such as wave dispersion \cite{You2021,lu2023nonparametric,fan2023bayesian,zhang2022metanor}. 
However, the above work relies on a predefined nonlocal model form and has only been applied to linear response prediction. 
Although there has been rapid progress on nonlocal models, to the best of our knowledge, there is no general procedure for determining the form of a nonlocal constitutive law for complex material behaviors, such as anisotropy, nonlinearity, etc.
The present work provides a novel method for addressing this.


In this paper, we aim to address the question of how to obtain from full-field spatial measurements a homogenized surrogate model  that captures complex material responses with minimum model-form error. 
To mitigate the three challenges discussed above encountered in the classical PDE parameter-fitting, and to exploit the advantages of nonlocal operators, we develop a novel nonlocal neural operator architecture. In contrast to existing model-free nonlocal neural operator-based material modeling approaches \cite{li2020neural,li2020fourier,you2022nonlocal,you2022learning,zhang2023metano} where the neural operator is constructed to find a solution operator mapping the body loads/boundary conditions to the corresponding displacement fields, our model uses nonlocal neural operators to find constitutive laws from data. To accomplish this, we leverage the nonlocal operator regression approach, to \textit{identify both the optimal nonlocal kernel function and the model form}. 
As summarized in Section~\ref{sec:peri}, a main innovation is that we parameterize the nonlocal neural operator as a peridynamic material model \cite{silling2000reformulation,silling2007peridynamic,madenci2016ordinary,bobaru2016handbook}. 
As such, the model exploits the expressivity of neural networks while it inherits from peridynamics adherance to the basic physics requirements such as the balance laws for linear and angular momentum, as well as objectivity. 
This architecture helps to capture the nonlinear homogenized material response as a function-to-function mapping. In Table \ref{tab:comparison} we summarize relevant properties of the proposed nonlocal constitutive law model in comparison with other state-of-the-art data-driven material models.

%

We summarize below our main contributions:
\begin{itemize}
\item We propose the Peridynamic Neural Operator (PNO), a new nonlocal neural operator-based formulation for capturing complex material responses. 
The PNO learns an optimal nonlocal constitutive law from full-field, spatial measurements of displacement and loading. 
In contrast to classical PDE-based models, our architecture learns the nonlocal bond forces as neural networks.
This captures nonlinear and anisotropic material responses without prior expert-constructed knowledge on the functional form of the strain energy density.
\item Because the neural operator design is an implementation of a special class of peridynamic material models
(``mobile'' materials), the PNO automatically satisfies the physically required balance laws and material objectivity. 
This implies that any simulation using PNO is compliant with these fundamental laws, even on samples outside the training range. This is a significant improvement over generic free-form neural networks.
\item By learning a constitutive model instead of a solution operator, our learned model is equally accurate for test samples with various resolutions, loading scenarios, and domain settings, even if they are substantially different from the ones used for training.
\item Our model learns directly from displacement field--loading field data pairs, 
which makes it particularly promising for learning complex material responses without explicit constitutive models and microstructure measurements. 
To demonstrate this capability, we provide in Section~\ref{sec:exp} examples of learning complex mechanical responses of materials directly from molecular dynamics and digital image correlation (DIC) tracking measurements.
\end{itemize}

\textbf{Outline of the paper.} Section \ref{sec:background} provides a brief introduction to the peridynamic nonlocal mechanical theory, including the class of material models provided by the PNO, and its connection with the graph-based nonlocal neural operators. 
In Section \ref{sec:pno} we describe the PNO architecture, together with theoretical results on its guaranteed physical consistency. 
To provide a condensed overview of the learning procedure for practitioners, a summary of the key ingredients and steps can be found in Algorithm \ref{alg:ML}. 
Section \ref{sec:exp} demonstrates the effectiveness of the learning technique on both synthetic and experimental datasets. 
We illustrate the model's generalizability with respect to loading, domain size, and domain shape, along with its expressivity with respect to nonlinear and anisotropic material responses. 
It is shown that in comparison 
with available conventional constitutive models, our method reduces the prediction error by around $30\%$ on a soft biological tissue data set based on DIC-tracked displacement measurements. 
Section \ref{sec:conclusion} summarizes our contributions and suggests future research directions.

\begin{table}
\begin{center}
{\small\begin{tabular}{ c | c | c | c | c | c |}
\hline
 Model &Generalizable& Generalizable  & Flexible  & Homogenized & Basic \\
       & in resolution          & in domain & model forms & effects& physical laws\\ \hline
Pre-defined models & \cmark & \cmark & -- & -- & \cmark\\
\hline
NODE-based constitutive laws& \cmark & \cmark & \halfcheckmark\footnotemark & -- & \cmark\\
\hline
Model-free neural networks& \halfcheckmark\footnotemark & -- & \cmark & -- & -- \\ \hline
Nonlocal operator regression & \cmark & \cmark & -- &  \cmark & \cmark \\ \hline
Peridynamic neural operators & \cmark & \cmark & \cmark & \cmark & \cmark \\ \hline
\end{tabular}}
\end{center}
\caption{List of properties for different experimental data-driven material-modeling approaches.}
\label{tab:comparison}
\end{table}

\addtocounter{footnote}{-1}
\footnotetext[1]{Since the strain energy density function is formulated as a neural network of invariants of the right Cauchy-Green deformation tensor \cite{tac2022data}, the model form depends on the pre-defined choice of invariants.}
\addtocounter{footnote}{1}
\footnotetext{When using nonlocal neural operator architectures, the model can be resolution-invariant \cite{li2020neural,li2020fourier}. Otherwise, the generic vector-to-vector mapping neural networks are not generalizable in measurement grids.}


\section{Background}\label{sec:background}

Throughout this paper, we use unbolded case letters to denote scalars/scalar-valued functions, bold letters to denote vectors/vector-valued functions, underlined unbolded letters for scalar-valued state functions, underlined bold letters for vector-valued state functions, and calligraphic letters for operators.
For any vector $\vb$, we use $\verti{\vb}$ to denote its $l^2$norm. For any function $\fb(\xb)$, $\xb\in\omg \subseteq\real^d$, taking values at nodes $\chi :=\{\xb_1,\xb_2,\dots,\xb_M\}$, $\vertii{\fb}$ denotes its $l^2$ norm, i.e., $\vertii{\fb}:=\sqrt{\sum_{i=1}^M {\verti{\fb(\xb_i)}^2}/M}$, which can be seen as an approximation to the $L^2(\omg)$ norm of $\fb$ (up to a constant). $\real^d$ represents the dimension-$d$ Euclidean space.

In this section, we briefly introduce the concept of peridynamic theory and nonlocal neural operators. Moreover, we review relevant concepts of conservation laws and two types of invariances, which will later become complementary to the proposed PNO definition.

\subsection{Peridynamic Theory}\label{sec:peri}

The peridynamic theory of solid mechanics (``peridynamics'') is an extension of the classical PDE theory 
that is compatible with discontinuous deformations, especially fractures.
In peridynamics, considering a domain of interest, $\omg\subset\real^d$, the equation of motion is given by
\begin{equation}
    \rho(\xb)\ddot\yb(\xb,t)= \int_{{B_\delta(\xb)}}\fb(\qb,\xb,t)\;d\qb+\bb(\xb,t)\text{ ,}\quad (\xb,t)\in\omg\times[0,T]\text{ ,}
\label{eqn-pdeomy}
\end{equation}
where $\xb$ and $\qb$ are material points in the reference (undeformed) configuration of the body
and $\yb$ is the deformation map.
$\rho(\xb)$ is the mass density function.
$B_\delta(\xb)$ is a ball centered at $\xb$ of radius ({\emph{horizon}}) $\delta$  that determines the extent of the long-range interactions. 
$\bb(\xb,t)$ is the body force density (external loading), which is assumed to be prescribed.
$\fb(\qb,\xb,t)$ is the {\emph{pairwise bond force density}}\footnote{There is not necessarily a direct physical interaction mechanism, like an electrostatic field, since the interaction in peridynamics can be indirect.} 
(dimensions of force per unit volume squared in 3D) that $\qb$ exerts on $\xb$. 
For any admissible pairwise bond force density field, it is required that
\begin{equation}
    \fb(\qb,\xb,t)=-\fb(\xb,\qb,t)
\label{eqn-fsym}
\end{equation}
for all $\xb$, $\qb$, and $t$,
which ensures that linear momentum is balanced globally, as discussed further below.

The material model determines $\fb(\qb,\xb,t)$ for every $\xb$ and $\qb$, for all $t$, and for every possible deformation.
To specify a material model, objects called {\emph{states}} are used.
A state is a mapping from a bond $\qb-\xb$ to some other quantity, usually either a vector or a scalar.
The notation for states is as follows:
\begin{equation}
  \ubA[\xb,t]\langle\xib\rangle = \bv\text{ ,}
\label{eqn-statenotation}
\end{equation}
where $\bv$ is the value of the state evaluated at the bond $\xib=\qb-\xb$.
The square brackets indicate that the state itself is defined at material point $\xb$ and time $t$. As two common examples of scalar-valued states, it is convenient to define:
\begin{equation}
  \ux\langle\xib\rangle=|\xib|\text{ ,} \qquad \uone\langle\xib\rangle=1\text{ ,} \qquad \text{for all bonds }\;\xib\text{ .}
\label{eqn-xstate}
\end{equation}
Two states that are used in material modeling are the {\emph{deformation state}} $\ubY$ and the {\emph{force state}} $\ubT$.
The deformation state maps bonds (which are defined in the reference configuration) to their images under the 
deformation:
\begin{equation}
  \ubY[\xb,t]\langle\qb-\xb\rangle=\yb(\qb,t)-\yb(\xb,t)\text{ .}
\label{eqn-Ydef}
\end{equation}
The force state contains the contribution of the material model at a point to the bond force density:
\begin{equation}
  \fb(\qb,\xb,t)=
  \ubT[\xb,t]\langle\qb-\xb\rangle-\ubT[\qb,t]\langle\xb-\qb\rangle\text{ .}
\label{eqn-Tdef}
\end{equation}
In \eqref{eqn-Tdef}, the terms on the right hand side are provided by the material model $\hat\ubT$ applied at
the points $\xb$ and at $\qb$:
\begin{equation}
   \ubT[\xb,t] =\hat\ubT(\ubY[\xb,t],\xb)\text{ ,} \qquad \ubT[\qb,t] =\hat\ubT(\ubY[\qb,t],\qb)\text{ .}
   \label{eqn-hatTdef}
\end{equation}
Thus, in peridynamics, a material model is a state-valued function of a state, rather than a tensor-valued
function of a tensor as in the PDE theory.
In a homogeneous body, there is no explicit dependence of the material model on $\xb$ or $\qb$, so that
\eqref{eqn-hatTdef} specializes to
\begin{equation}
   \ubT[\xb,t] =\hat\ubT(\ubY[\xb,t])\text{ ,} \qquad \ubT[\qb,t] =\hat\ubT(\ubY[\qb,t])\text{ .}
   \label{eqn-hatThomog}
\end{equation}
In all previous applications of peridynamics known to the authors, the material model is given by an algebraic expression, for
example
\begin{equation}
  \ubT=3\ue\,\ubM \quad{\mathrm{or}}\quad
   \ubT\langle\xib\rangle=3\ue\langle\xib\rangle\ubM\langle\xib\rangle\text{ ,}
\label{eqn-Texample}
\end{equation}
where $\xib:=\qb-\xb$ is a generic bond vector and $\etab:=\ub(\qb,t)-\ub(\xb,t)$ is the generic relative displacement between the bond endpoints. Then we can formulate the deformation state $\ubY$ as
\begin{equation}
  \ubY[\xb,t]\langle\qb-\xb\rangle = \xib+\etab\text{ .}
  \label{eqn-Yeta}
\end{equation}
In \eqref{eqn-Texample}, $\ubM$ is a state that contains unit vectors in the direction
of the deformed bonds:
\begin{equation}
  \ubM\langle\xib\rangle=\frac{\ubY\langle\xib\rangle}{|\ubY\langle\xib\rangle|}=\frac{\xib+\etab}{|\xib+\etab|}\text{ ,}
\label{eqn-Mdef}
\end{equation}
and the {\emph{extension state}} $\ue$ is a scalar valued state that contains the length changes of the
bonds as the body deforms:
\begin{equation}
   \ue\langle\xib\rangle=|\ubY\langle\xib\rangle|-|\xib|= |\xib+\etab|-|\xib|\text{ .}
\label{eqn-extendef}
\end{equation}
In contrast to this usual practice of using an algebraic expression for the material model, in the present work, 
the material model is provided by a nonlocal neural operator.
As discussed further below,
the mathematical structure of the implementation ensures its physical admissibility.

All the material models developed in the present work are {\emph{ordinary}}, meaning that the bond force vectors
in the force state are always parallel to the deformed bonds.
Thus, there is a scalar-valued state $\ut$ such that
\begin{equation}
  \hat\ubT(\ubY) = \ut(\ubY)\,\ubM \quad{\mathrm{or}}\quad
  \ubT\langle\xib\rangle=\ut\langle\xib\rangle \ubM\langle\xib\rangle \qquad \text{for all bonds }\;\xib\text{ .}
\label{eqn-ord}
\end{equation}
All the material models in this paper are also {\emph{mobile}}, which means that the scalar force state
in \eqref{eqn-ord} depends only on the length changes of the bonds\footnote{The term ``mobile'' was coined because of the slight resemblance of these materials to works of art that
are suspended from the ceiling and have different parts that rotate freely,
independent of each other.}:
\begin{equation}
 \hat\ubT(\ubY) = \ut(\ue)\,\ubM\text{ ,}
\label{eqn-mobile}
\end{equation}
where $\ue$ is given by \eqref{eqn-extendef}.
Clearly, mobile implies ordinary. 

Combining \eqref{eqn-pdeomy}, \eqref{eqn-Tdef}, and \eqref{eqn-ord}, we obtain the following  governing equation 
\begin{equation}\label{eqn:peri_full}
\rho(\xb)\ddot\yb(\xb,t)=\int_{{B_\delta(\mathbf{0})}}\left(\ut[\xb,t]\langle\xib\rangle+\ut[\xb+\xib,t]\langle -\xib\rangle\right) \ubM[\xb,t]\langle\xib\rangle\;d\xib +\bb(\xb,t),
\end{equation}
where the identity $\ubM[\xb+\xib,t]\langle-\xib\rangle=-\ubM[\xb,t]\langle\xib\rangle$ has been used.

Now consider the status of peridynamics with respect to the basic laws of physics.
\begin{itemize}
\item[$\bullet$] {\emph{Conservation of linear momentum.}}
First, observe that a pairwise force density $\fb$ of the form \eqref{eqn-Tdef} necessarily satisfies
the antisymmetry requirement \eqref{eqn-fsym}.
It is shown (Proposition 7.1 of \cite{silling2007peridynamic}) that if this antisymmetry
condition holds, then the global balance of
linear momentum holds in a bounded body $\calB$:
\begin{equation}
  \eqref{eqn-fsym} \quad\implies\quad
  \frac{d}{dt}\int_\calB \rho(\xb)\dot\yb(\xb,t)\;d\xb = \int_\calB \bb(\xb,t)\;d\xb\text{ .}
\label{eqn-global-lin}
\end{equation}
Mechanically, the antisymmetry condition means that each endpoint of any bond experiences a
force vector that is equal in magnitude and opposite in direction to the other, so the result
\eqref{eqn-global-lin} is very reasonable.
\item[$\bullet$] {\emph{Conservation of angular momentum.}}
It is shown that a body composed of an ordinary material necessarily satisfies the
global balance of angular momentum (Proposition 8.2 of \cite{silling2007peridynamic}).
That is,
\begin{equation}
  \eqref{eqn-ord} \quad\implies\quad
  \frac{d}{dt}\int_\calB \rho(\xb)\yb(\xb,t)\times\dot\yb(\xb,t)\;d\xb = \int_\calB\yb(\xb,t)\times\bb(\xb,t)\;d\xb\text{ .}
\label{eqn-global-ang}
\end{equation}
The mechanical picture in an ordinary material is that since the bond force is parallel to the
bond direction, each endpoint of a bond exerts no net moment on the other, thus ensuring the balance
of angular momentum.
\end{itemize}

The following two invariances do not have the same status as the conservation laws but are
required for physical reasonableness:
\begin{itemize}
\item [$\bullet$] {\emph{Galilean invariance.}}
This requirement, also called translational invariance, means that if a body is rigidly translated,
its response to deformation remains unchanged.
This invariance is built into the basic structure of peridynamic material modeling.
To see this, recall that according to \eqref{eqn-hatTdef}, the bond forces depend only on $\ubY$.
If the body is translated through a constant vector $\Cb\in\real^d$, then from 
\eqref{eqn-Ydef},
\begin{equation}
  \ubY[\xb,t]\langle\qb-\xb\rangle=\big(\Cb+\yb(\qb,t)\big)-\big(\Cb+\yb(\xb,t)\big)=
  \yb(\qb,t)-\yb(\xb,t)\text{,}
\label{eqn-Ytranslate}
\end{equation}
leaving $\ubY$ unchanged.
The translation does not change $\xb$ and $\qb$, which are defined in the reference configuration.
Thus, any peridynamic material model necessarily satisfies Galilean invariance without additional conditions.
\item [$\bullet$] {\emph{Objectivity.}}
Objectivity, also called material frame invariance, is satisfied by a material model if for any proper orthogonal tensor $\Rb\in\real^{d\times d}$,
\begin{equation}
  \hat\ubT(\ubY')=\Rb\hat\ubT(\ubY)\text{ ,}
  \qquad{\mathrm{where}}\qquad
  \ubY'\langle\xib\rangle=\Rb\ubY\langle\xib\rangle\text{ .}
\label{eqn-objdef}
\end{equation}
Mechanically, objectivity means that if a body is rigidly rotated {\emph{after}} it is deformed,
then the bond force vectors can be found by rotating the bond force vectors that would be present in the unrotated deformation.
Objectivity should not be confused with isotropy, which pertains to what happens if the body is
rotated {\emph{before}} deforming it.
It is proved in Proposition 9.3 of \cite{silling2010peridynamic} that every mobile material model
satisfies objectivity, that is,
$\eqref{eqn-mobile}\implies\eqref{eqn-objdef}$.
\end{itemize}


Many material models have been developed for peridynamics, and any material model from the local theory can be translated into peridynamic form \cite{silling2010peridynamic,Tian2020,madenci2016peridynamic,foster2011energy}.
The most widely used capability that peridynamics offers that is not available in the local theory is the direct modeling of fracture within the basic field equations, since the equation of motion \eqref{eqn-pdeomy} is an integro-differential equation that does not involve the partial derivatives of displacement with respect to position \cite{diehl2019review,lipton2014dynamic,diehl2022comparative,ha2010studies,madenci2013peridynamic}. 
However, although the present paper concerns both infinitesimal and finite deformations of material responses, we do not address fracture. The extension of the methods described here to include fracture modeling is under investigation in separate work.

The peridynamic neural operator, described in the next Section, provides an ordinary, mobile material model.
It is helpful to first introduce the general structure of this type of model.
A typical ordinary material model is the linear peridynamic solid (LPS) \cite{silling2007peridynamic}, which
is similar to an isotropic linear elastic solid in the PDE theory.
The LPS can be written in the following form:
\begin{equation}
 \ut\langle\xib\rangle=C_1\vartheta d\,\uomega\langle\xib\rangle\,\ux\langle\xib\rangle + C_2\,\uomega\langle\xib\rangle\,\ue\langle\xib\rangle=\uomega\langle\xib\rangle\,\left(C_1\vartheta d\,\ux\langle\xib\rangle + C_2\,\ue\langle\xib\rangle\right)\text{ .}
\label{eqn-lps}
\end{equation}
Here the {\emph{nonlocal dilatation}} is defined by
\begin{equation}
 \vartheta=\dfrac{d\int_{B_\delta(\mathbf{0})}\uomega\langle\xib\rangle\, \ue\langle\xib\rangle\verti{\xib}d\xib }{\int_{B_\delta(\mathbf{0})}\uomega\langle\xib\rangle\, \verti{\xib}^2d\xib}\text{ .}
\label{eqn-thetadef}
\end{equation}
In \eqref{eqn-lps}, $C_1$ and $C_2$ are constants that are chosen so that the model reproduces
the correct bulk modulus and shear modulus, and 
$\uomega$ is a weighting function called the {\emph{influence state}}. 
In an isotropic expansion of a body with a small volume change, the nonlocal dilatation defined by  \eqref{eqn-thetadef}
closely approximates the usual dilatation from the PDE theory (the trace of the strain tensor). An important point is that in the LPS, although $\vartheta$ depends on the deformation of all the bonds,
this dependence appears only through the changes in bond length.
Thus the LPS is consistent with \eqref{eqn-mobile} as well as \eqref{eqn-ord}, so it is
both ordinary and mobile.


In general, the functional form of $\ut$ and $\uomega$ as they depend on $\xib$ cannot be established a priori
for a given material.
The integral constitutive laws incorporate the nonlinear and heterogeneous responses of the system and must be chosen to reproduce these possibly complex responses separately for each material. 
To accomplish this, our proposed method learns $\ut$ and $\uomega$ from data using neural networks. As such, we obtain a data-driven nonlocal constitutive law in the form of a neural operator architecture, as described in Section \ref{sec:pno}.

\subsection{Nonlocal Neural Operators}

We now introduce the basic concepts and architectures of the general nonlocal neural operators \citep{li2020neural,li2020multipole,li2020fourier,you2022nonlocal,you2022learning}, which were developed for scientific computing applications entailing the learning of solution operators. A prototypical instance is the case of solving the governing PDE in material modeling problems, where the initial input field (body load/boundary load) is mapped to the corresponding displacement field via a nonlinear parameterized mapping. The demand for operator learning has sparked the development of several architectures in neural operator based methods \citep{li2020neural,li2020multipole,li2020fourier,you2022nonlocal,you2022physics,you2022learning,goswami2022physics,liu2023ino,gupta2021multiwavelet,lu2019deeponet,cao2021choose,yin2022continuous,hao2023gnot,li2022transformer,yin2022continuous}. Contrary to classical NNs that operate between finite-dimensional Euclidean spaces, neural operators are designed to learn mappings between infinite-dimensional function spaces \citep{li2020neural,li2020multipole,li2020fourier,you2022nonlocal,Ong2022,gupta2021multiwaveletbased,lu2019deeponet,lu2021learning,goswami2022physics, gupta2021multiwavelet}. A remarkable advantage of neural operators lies in their resolution independence, which implies that the prediction accuracy is invariant to the resolution of input functions. Furthermore, in contrast to classical PDE-based approaches, neural operators can be trained directly from data, and hence requires no domain knowledge nor pre-assumed PDEs. All these advantages make neural operators a promising tool for learning complex material responses \citep{yin2022simulating,goswami2022physics,yin2022interfacing,you2022physics,li2020neural,li2020multipole,li2020fourier,lu2021comprehensive}.

Herein, we focus on the nonlocal neural operator introduced in \cite{li2020neural}, which has foundation in the representation of the solution of a PDE by the Green's function. For the purpose of introduction, we focus on the static PDE solving scenario, which aims to construct a surrogate operator $\mcG:\mbF\rightarrow \mbU$ that maps the input function $\bb(\xb)$ to the output function $\ub(\xb)$. In the context of material modeling \cite{you2022learning,you2022physics}, $\bb$ stands for the boundary and body loads, and $\ub$ is the corresponding displacement or damage fields. 
This architecture is comprised of three building blocks. First, the input function, $\bb(\xb)\in\mbF$, is lifted to a higher-dimension representation via a local affine pointwise mapping $\hb(\xb,0)=\mathcal{P}[\bb](\xb)$. Next, the feature vector function $\hb(\xb,0)$ goes through an iterative layer block where the layer update is defined via nonlinear operator layers:  $\hb(\cdot,l+1)=\mathcal{J}[\hb(\cdot,l)]$, for $l=0,\cdots,L-1$, which will be further discussed in the later contents. Here, $\hb(\cdot,l)$ is the $l-$th layer network feature, which is a dimension-$d_h$ vector-valued function defined on $\omg$. Finally, the output $\ub(\cdot)\in\mbU$ is obtained through a projection layer: $\ub(\xb)=\mathcal{Q}[\hb(\cdot,L)](\xb)$. 

In nonlocal neural operators, the resolution-independence property is realized by parameterizing the layer update, $\mcJ$, as an integral operator. In particular, for an $L-$layer NN, the $l-$th layer network update is given as:
\begin{equation}\label{eq:gkn}
\hb(\xb,l+1)=\mathcal{J}[\hb(\cdot,l)](\xb):=\sigma\left(\Rb\hb(\xb,l)+\int_\omg \Kb(\xb,\qb,\bb(\xb),\bb(\qb);\vb)\hb(\qb,l) d\qb + \cb\right)\text{ .}
\end{equation}
Here, $\sigma$ is an activation function, $\Rb\in\real^{d_h\times d_h}$ is a trainable tensor, $\cb\in\real^{d_h}$ is a trainable vector, and $\Kb\in\real^{d_h\times d_h}$ is a tensor kernel function that takes the form of a (usually shallow) NN whose parameters $\vb$ are to be learned. While in the original version of nonlocal neural operator the integral is extended to the whole set $\omg$, for efficiency purposes, restrictions to a ball of radius $\delta$ centered at $\xb$, i.e. $B_\delta(\xb)$, can also be considered. However, as expected, this choice might compromise the accuracy since the support of Green's function generally spans the whole domain in PDE solving problems \cite{li2020neural}.

The implementation of nonlocal neural operators \eqref{eq:gkn} resembles graph neural networks in that both follow a message passing framework \citep{gilmer2017neural,brandstetter2021message}, while the discrete aggregation step is substituted by a meshfree approximation of the continuous integral operator \cite{liu2023harnessing}. In particular, taking a discretization of $\ub$ on a collection of points $\chi=\{\xb_j\}^M_{j=1}\subset \omg$, the kernel integration can be viewed as an aggregation of messages by taking the edge feature:
\begin{equation}
\eb(\xb_i,\xb_j):=\Kb(\xb_i,\xb_j,\bb(\xb_i),\bb(\xb_j);\vb)\text{ ,}
\end{equation}
and the node feature update:
\begin{equation}
\hb(\xb_i,l+1)=\sigma\left(\Rb\hb(\xb_i,l)+\sum_{x_j\in N(\xb_i)}\eb(\xb_i,\xb_j)\hb(\xb_j,l)\mu(\xb_j) + \cb\right)\text{ ,}
\end{equation}
where $N(\xb_i)$ is the neighborhood of $\xb_i$ according to the graph on  $\chi$ and $\mu(\xb_j)$ is the Riemannian sum weight. Since this nonlocal layer update is consistent with a continuous integral operator, the learned network parameters are resolution-independent: 
the learned $\vb$, $\Rb$, and $\cb$ can be reused even when the grid set, $\chi$, varies, since they correspond to the same integral operator $\mcJ$. Therefore, the nonlocal architecture provides the key for resolution independence.

Despite the aforementioned advances of neural operators, purely data-driven neural operators need to learn the basic physical laws from data, and therefore suffer from data challenge. In particular, in order to generalize the solution, they require a large corpus of paired datasets, which is prohibitively expensive in many engineering applications. To resolve this challenge, the physics-informed neural operator (PINO) \cite{li2021physics} and physics-informed DeepONets \cite{goswami2022physics,wang2021learning} are introduced, where a PDE-based loss is added to the training loss as a penalization term. However, these approaches still require \textit{a priori} knowledge of the underlying PDEs, which restricts their applications to (known) constitutive law-solving tasks. As another restriction, since most neural operator model learns the material responses as a surrogate solution operator, they are not generalizable to different domain geometries and loading scenarios. That means, an operator learned on square-domain samples cannot be immediately applied to predict the material deformation on a circular domain. To resolve these two limitations, in this work we parameterize the nonlocal neural operator architecture as a nonlocal constitutive law based on the ordinary state-based peridynamics formulation \eqref{eqn-ord}. As such, the model preserves the fundamental physical laws and is generalizable to different domain geometries and loading scenarios.

\section{Peridynamic Neural Operators}\label{sec:pno}

\subsection{Proposed Architecture}\label{sec:pno_formulation}

In this Section, we introduce the Peridynamic Neural Operator (PNO) framework, which learns ordinary state-based peridynamic constitutive models from data. 
It uses the nonlocal neural operator architecture \cite{li2020neural,you2022nonlocal}. 
The PNO is used as the surrogate for the integrand functions in \eqref{eqn-pdeomy}.
The model is designed to represent diverse features of material behaviors, such as anisotropy and nonlinearity. By taking the form of \eqref{eqn-ord} and \eqref{eqn-mobile}, the proposed operator inherits the physical compatibility 
of the mobile ordinary state-based materials within the peridynamic theory of mechanics.
This guarantees the required invariances under frame translation and rotation, as well as
the balances of linear and angular momentum.

In the following, it is convenient to use displacement rather than the deformation map directly.
These are related by
\begin{equation}
  \ub(\xb,t)=\yb(\xb,t)-\xb\text{ .}
  \label{eqn-udef}
\end{equation}
Formally, we attempt the learning of complex physical responses in a mechanical system, based on a set of observations $\mcD=\{\bb^s,\ub^s\}_{s=1}^S$ of the loading field $\bb^s(\xb,t)$ 
and the corresponding displacement field $\ub^s(\xb,t)$, where 
$s$ is the sample index. 
Inspired by the peridynamic formulation \eqref{eqn:peri_full}, we propose to learn a nonlocal neural operator $\mcG$, which acts as a surrogate operator for the constitutive law such that:
\begin{equation}
  \mcG[\ub^s](\xb,t)\approx \rho(\xb)\ddot\ub^s(\xb,t)-\bb^s(\xb,t)\text{ ,}
  \label{eqn-Gapprox}
\end{equation}
where the operator $\mcG$ is formulated as:
\[\mcG[\ub](\xb,t):=\int_{{B_\delta(\mathbf{0})}}\left(\ut[\xb,t]\langle\xib\rangle+\ut[\xb+\xib,t]\langle -\xib\rangle\right) \ubM[\xb,t]\langle\xib\rangle\;d\xib.\]
Herein, we parameterize the scalar force state and the influence state with neural networks: \begin{equation}\label{eqn:umt}
  \ut[\xb,t]\langle{\xib}\rangle:= \sigma^{NN}(\omega(\xib), \vartheta(\xb,t), e(\xib,\etab), |\xib|;\vb)\text{ ,}
\end{equation}
and
\begin{equation}\label{eqn:omega_aniso}
  \omega(\xib) :=\omega^{NN}(\xib;\wb)\text{ ,}
\end{equation}
where $\sigma^{NN}$ and $\omega^{NN}$
are scalar-valued functions that take the form of a (usually shallow) multi-layer perceptron (MLP) with learnable parameters $\vb$ and $\wb$, respectively. $e(\xib,\etab):=\verti{\xib+\etab}-\verti{\xib}$.  
As a generalization of the nonlocal dilatation defined in \eqref{eqn-thetadef}, we set 
\begin{equation}\label{eqn:dila}
  \vartheta(\xb,t) :=\dfrac{\int_{B_\delta (\mathbf{0})} \omega^{NN}\left(\xib;\wb\right)e(\xib,\etab)|\xib| d \xib}{\int_{B_\delta (\mathbf{0})} \omega^{NN}\left(\xib;\wb\right)|\xib|^2 d \xib}\text{ .}
\end{equation}
If  the material is   known to be isotropic a-priori, the following simplification of \eqref{eqn:omega_aniso} holds:
\begin{equation}\label{eqn:omega_iso}
  \omega(\xib) :=\omega^{NN}(|\xib|;\wb).
\end{equation} 

Given a forcing term $\bb^s$, in order to guarantee the existence of a unique solution $\ub^s$, 
nonlocal boundary conditions (``volume constraints'') must be prescribed on an appropriate {\it interaction domain} $\Omega_I$. 
Without loss of generality, the Dirichlet boundary condition is applied, with
$\Omega_I=\{\xb|\xb\in\real^d \backslash \omg,\,\text{dist}(\xb,\omg)<2\delta\}$.
As will be elaborated in our experiments of Section \ref{sec:exp}, the measurements of the displacement field $\ub^s$ are defined either with a periodic boundary condition or properly extended using mirror boundary conditions \cite{foss2023convergence}, and therefore they are available on $\overline{\omg}:=\omg\cup\omg_I$.

Since the layer update in the above PNO is formulated as a continuous integral operator, the learned network parameters are resolution-independent: the learned $\vb$ and $\wb$ are close to optimal even when used with different resolutions.
Importantly, since it is implemented in the form of an ordinary, mobile material model,
the PNO preserves the balance of linear and angular momentum and satisfies objectivity, as discussed previously. Similar to the vanilla nonlocal neural operators, PNO can be implemented using a two-layer message passing framework. In particular, for each grid point $\xb_i\in\chi$ and time instance $t$, we calculate the first edge feature in the vanilla nonlocal neural operator as:
\begin{equation}
\eb^1(\xb_i,\xb_j):=\omega^{NN}(\xb_j-\xb_i;\wb)\text{ ,}
\end{equation}
and use it to update the first node feature: 
\begin{equation}
\nb^1(\xb_i,t):=\dfrac{\sum_{N(\xb_i)}\eb^1(\xb_i,\xb_j)(\verti{\xb_j+\ub(\xb_j,t)-\xb_i-\ub(\xb_i,t)}-\verti{\xb_j-\xb_i})|\xb_j-\xb_i| \mu(\xb_j)}{\sum_{N(\xb_i)}\eb^1(\xb_i,\xb_j)|\xb_j-\xb_i|^2 \mu(\xb_j)}\text{ .}
\end{equation}
Then, as the second layer, we update the second edge feature via:
\begin{equation}
\eb^2(\xb_i,\xb_j,t):=\sigma^{NN}(\eb^1(\xb_i,\xb_j),\nb^1(\xb_i,t),\verti{\xb_j+\ub(\xb_j,t)-\xb_i-\ub(\xb_i,t)}-\verti{\xb_j-\xb_i},|\xb_j-\xb_i|;\vb)\text{ ,}
\end{equation}
and finally the second node feature as the operator output:
\begin{equation}
\mcG[\ub](\xb_i,t)=\nb^2(\xb_i,t):=\sum_{N(\xb_i)}\left(\eb^2(\xb_i,\xb_j,t)+\eb^2(\xb_j,\xb_i,t)\right)\dfrac{\xb_j+\ub(\xb_j,t)-\xb_i-\ub(\xb_i,t)}{\verti{\xb_j+\ub(\xb_j,t)-\xb_i-\ub(\xb_i,t)}}\mu(\xb_j)\text{ .}
\end{equation}
Therefore, our PNOs share similar properties to the peridynamic models in its ability to preserve basic physical properties, and can also be seen as a special architecture of nonlocal neural operators.

\subsection{Learning Algorithm}\label{sec:alg}

\begin{algorithm}
\caption{Algorithm for PNOs.}\label{alg:ML}
\begin{algorithmic}[1]
\State \textbf{Pre-processing Phase:}
\State Read data and inputs: $\chi=\{\xb_i\}_{i=1}^N$, $\mcD_{tr}:=\{\ub^{s,tr}_i,\bb^{s,tr}_i\}_{i=1,s=1}^{N,S_{tr}}$ for training, $\mcD_{val}:=\{\ub^{s,val}_i,\bb^{s,val}_i\}_{i=1,s=1}^{N,S_{val}}$ for validation sets.
\State Find and store the peridynamic neighbor node set $N(\xb_i)$ in the form of message passing graph structure:
\For{$i = 1:N$}
    \State Find all $\xb_j$ such that $|\xb_j - \xb_i|< \delta$.
\EndFor
\State \textbf{Training Phase:}
\For{$ep = 1: epoch_{max}$}
    \State Initialize error: $E_{tr} = 0$
    \For{ each batch}
        \State Reset batch loss: $loss = 0$
        \If{Case I}
            \State Compute batch loss following \eqref{eqn: caseI_loss}.
        \ElsIf{Case II}
            \State For each training sample in this batch, run nonlinear solver to obtain $\mcG^{-1}[\bb^s]$.
            \State Compute batch loss following \eqref{eqn: caseII_loss}.
        \EndIf
        \State Update PNO parameters $\wb$, $\vb$ with the Adam optimizer.
    \EndFor
    \State Compute training error $E_{tr}$.
    \State \textbf{Validation Phase:}
    \If{training error decreases}
        \State Initialize validation error: $E_{val} = 0$
        \If{Case I}
            \State Compute validation error following \eqref{eqn: caseI_loss}.
        \ElsIf{Case II}
            \State For each validation sample, run nonlinear solver to obtain $\mcG^{-1}[\bb^s]$.
            \State Compute validation error following \eqref{eqn: caseII_loss}.
        \EndIf
        \State If validation error also decreases, save the current model. 
    \EndIf
\EndFor
\end{algorithmic}
\end{algorithm}

In this Section we elaborate the learning algorithm of PNOs. Although the prescribed architecture is readily applicable to higher-dimensional domains and dynamic cases, in this work we focus on (quasi)static and two-dimensional ($d=2$) tasks. In this case, our goal is to learn the surrogate operator $\mcG$ such that
$$\mcG[\ub](\xb)\approx -\bb(\xb)\text{ ,}$$
Algorithm \ref{alg:ML} lays out the structure of the pseudo code used for this work to learn the consitutive model, and we will explain more details below. 
The program is coded in Python and employs PyTorch libraries for GPU computing. 
The code uses the Adam optimizer in PyTorch for minimizing the loss function.

Formally, we assume that measurements are available in the following format:
\begin{itemize}
    \item Coordinates of nodes where measurements are taken: $\chi=\{\xb_i\}_{i=1}^N$, where $N$ is the total number of nodes.
    \item External force-displacement function pairs at each node $\xb_i$ and for each sample $s$:  
    $\{\ub^{s}_i,\bb^s_i\}_{i=1,s=1}^{N,S}$ where $S$ is the total number of samples.
\end{itemize}
Moreover, we split the available measurements into three data sets, and let $S_{tr}$, $S_{val}$, $S_{test}$ denote the total number of samples in training, validation, and test sets, respectively.

The neural network functions for $\omega^{NN}$ and $\sigma^{NN}$ are multi-layer perceptrons (MLP) with three layers and ReLu as the activation function. 
The widths of the MLP layers are denoted by $(W_{in}, W_{h1}, W_{h2}, W_{out})$, where $W_{in}$ is the input's dimension, $W_{h1}, W_{h2}$ are the widths of the hidden layers, and $W_{out}$ is the dimension of the output. 
For $\omega^{NN}$, depending on the assumption of isotropy or anisotropy through (\ref{eqn:omega_iso}) or (\ref{eqn:omega_aniso}), 
$W_{in}$ can be 1 or 2. 
In this study, $W_{out}=1$ for all examples, although it can take on larger values in other applications. From (\ref{eqn:umt}), since $|\xib|$ and $e(\xib,\etab)$ are scalars and $\vartheta(\xb)$ has the same dimension as $\omega(\xib)$ (see (\ref{eqn:dila})), 
it follows that $W_{in}=4$ and $W_{out}=1$ in the $\sigma^{NN}$ MLP. 
$W_{h1}, W_{h2}$ in both $\omega^{NN}$ and $\sigma^{NN}$ as well as the horizon size $\delta$ are hyperparameters of choice to tune.
The larger these values are, the more expressive (and more expensive) the PNO becomes.

Depending on whether the data includes nonzero external forces (Case I) or not (Case II), 
either of the following two different loss functions is used. 
In Case I, we follow the convention of nonlocal neural operator literatures \cite{li2020neural,li2020fourier} and consider the relative $L^2$ error of the output function, $\bb$, as the loss function, which is given by
\begin{equation}\label{eqn: caseI_loss}
    \text{loss}_b = \frac{1}{S_{tr}}\sum_{s = 1}^{S_{tr}}\dfrac{\vertii{\mcG[\ub^s]+\bb^s}}{\vertii{\bb^s}}\text{ .} 
\end{equation}
Physically, the loss function in \eqref{eqn: caseI_loss} measures the error in the internal force distribution on the nodes needed to equilibrate the system.

If external forces are zero (Case II), the loss function \eqref{eqn: caseI_loss} fails in that the denominator becomes zero.
If the denominator is removed to avoid division by zero, the program leads to the trivial solution for the model parameters, 
such that with any deformation the model returns zero forces. 
To circumvent this problem, in Case II, we define the loss in terms of the displacement field:
\begin{equation}\label{eqn: caseII_loss}
   loss_u = \frac{1}{S_{tr}}\sum_{s = 1}^{S_{tr}}\dfrac{\vertii{\mcG^{-1}[\bb^s]-\ub^s}}{\vertii{\ub^s}}\text{ ,}
\end{equation}
where $\mcG^{-1}[\bb^s]$ denotes the numerical solution of $\mcG[\ub]=-\bb^s$ using an iterative nonlinear static solver (see line 21 in Algorithm~\ref{alg:ML}). In this work we use the Polak-Ribiere conjugate gradient method \cite{shewchuk1994introduction}, due to its efficiency and robustness in peridynamics nonlinear problems \cite{van2014relationship}. 
Additional inputs needed for the solver are ${\ub^{s}_0}$, ${\ub^{s}_{BC}}$, $tol$, $itr_{max}$, which denote the initialization displacement (which may or may not be zero), Dirichlet boundary condition values on $\Omega_I$, convergence tolerance, and maximum allowed iterations, respectively.

Two difficulties arise in Case II.
Firstly, the computational time significantly increases due to the multi-iteration procedure per sample and per epoch. 
Secondly and more importantly, each static-solver iteration (which includes at least one model evaluation) is technically a part of the loss function in the sense that it participates in auto-differentiation/back-propagation by the optimizer. 
This procedure requires a relatively large memory allocation compared with the Case I where the loss function is defined for forces and the model is evaluated just once for each sample in each epoch. 
Consequently, in Case II, the memory allocation strongly depends on the number of iterations the nonlinear solver converges in. 
To make the Case II algorithm affordable, considering hardware limitations, one can take the following measures to reduce the number of solver's iterations:
\begin{enumerate}
    \item Avoiding the use of too small tolerances for convergence criterion.
    \item Smart initialization of the solution. Examples of smart initialization values include using displacements from another sample with a close measurement sequence, or using some form of interpolation of Dirichlet boundary conditions. 
    \item Limiting the maximum number of iterations to the lowest possible value at which convergence for most samples is likely to be achieved.
\end{enumerate}
In Case I, a substantial number of different external loading fields can be used to train the PNO without excessive cost.
However, in Case II, the training requires a full peridynamic solution for each instance of the training data, 
which greatly increases the cost.
In Case II, we found it helps to modify the architecture of the scalar force state given by \eqref{eqn:umt} as follows:
\begin{equation}
    \ut[\xb]\langle{\xib}\rangle:= \sigma^{NN}(\omega(\xib), \vartheta(\xb), e(\xib, \etab), |\xib|;v) \; e(\xib, \etab)\text{ .} 
    \label{eqn:umtcaseii}
\end{equation}
The new factor of $e(\xib,\etab)$ guarantees that a small deformation bond will be associated with a small force state, and it causes the larger deformation bonds to weigh more on the overall equilibrium deformation. Thus, the modified form \eqref{eqn:umtcaseii} tends 
to improve the training efficiency. 

In the next section, the PNO is used to learn constitutive models for three mechanically different materials. The first two are examples of Case I, and the third example is of Case II.

\section{Numerical Examples}\label{sec:exp}

In this section, the PNO is applied to learn the constitutive laws for three different materials with distinct characteristics: 
\begin{enumerate}
    \item A single layer graphene at zero temperature.
    \item An anisotropic hyperelastic material going under large deformation.
    \item  A heterogeneous anisotropic biological tissue under large deformation.
\end{enumerate}
The first example is chosen to demonstrate that the PNO is capable of learning a continuum constitutive law from microscale data, in this case using molecular dynamics (MD). 
The second example illustrates the ability of the method to capture anisotropy and material response under large deformation. 
While the first two examples are based on synthetic data sets, the third example uses real-world experimental data from a tissue specimen with heterogeneity and complex behaviors, and the measurements also contain unavoidable noises \cite{you2022learning}. 
All tests are performed on a NVIDIA RTX 3090 GPU card with 24GB memory.

\subsection{Example I: Graphene}

\begin{figure}[!t]\centering
\includegraphics[width=.7\linewidth]{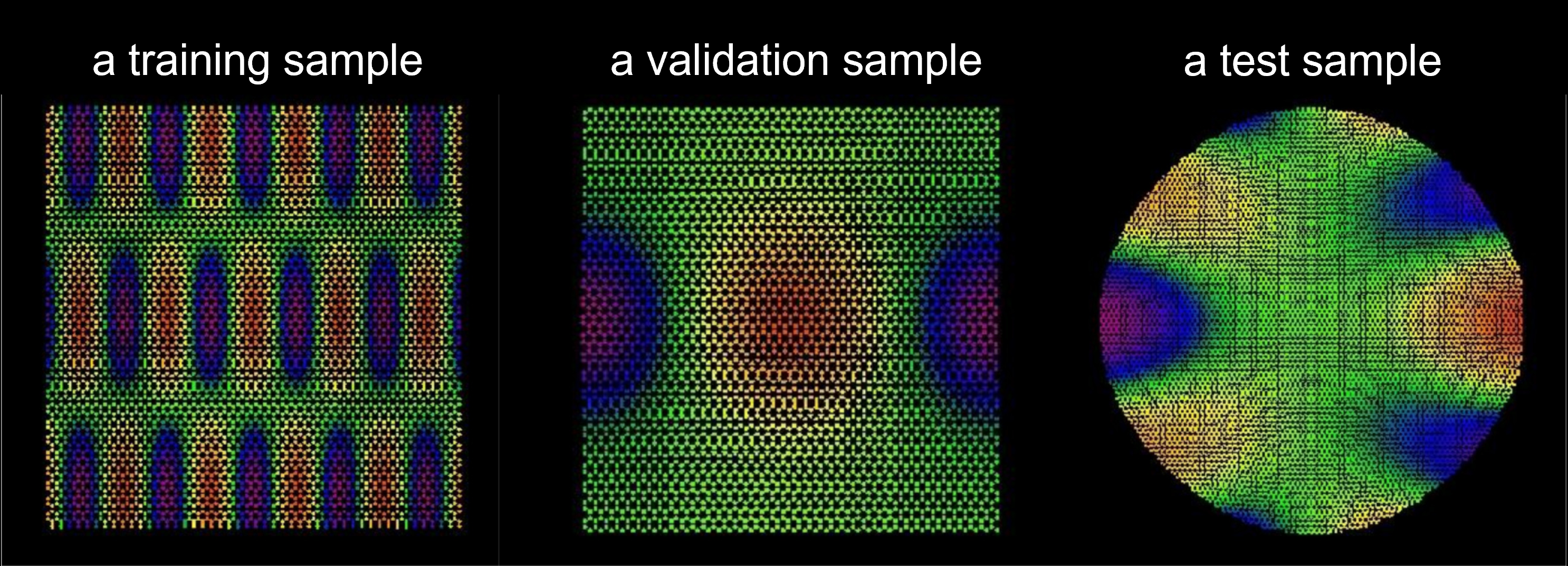}
 \caption{Demonstration of typical training, validation, and testing samples from MD data sets in Example I: Graphene.
 Colors show the $u_1$ displacement component.}
 \label{fig:MD_samples}
\end{figure}

\begin{figure}[!t]\centering
\includegraphics[width=.8\linewidth]{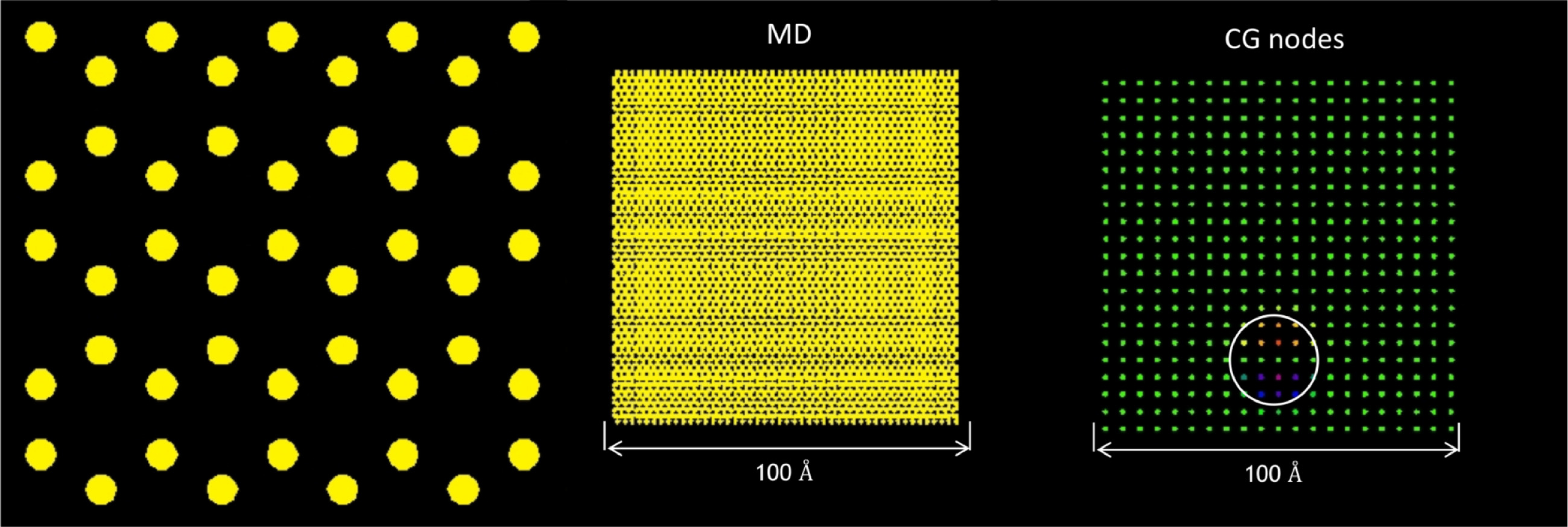}
 \caption{Demonstration of data generation for Example I: Graphene. Left: hexagongal atomic lattice in MD. Center: full MD mesh. Right: 21x21 nodes coarse-grained nodes.
 Colors within the small circle show coarse-grained bond forces on the node at the center of the circle.}
 \label{fig:MD_CG}
\end{figure}

\begin{figure}[!t]\centering
\includegraphics[width=.99\linewidth]{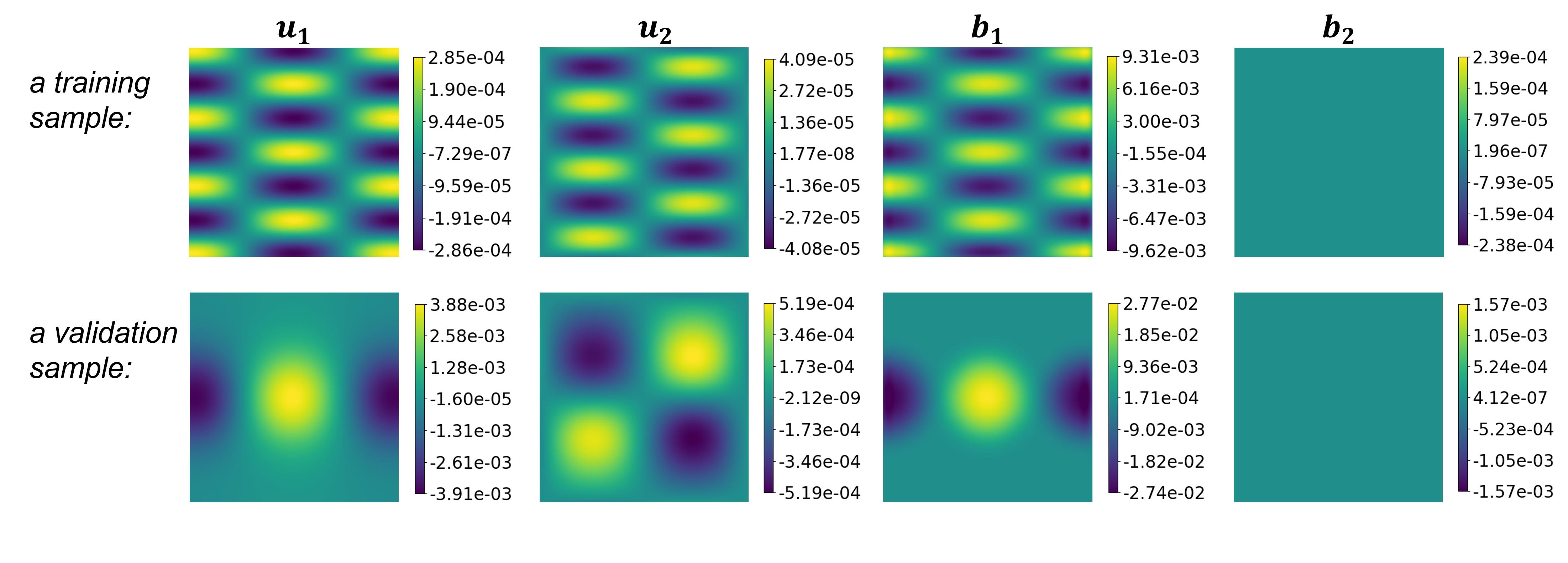}
 \caption{The body force and the corresponding displacement field components for a training and a validation sample of the coarse-grained graphene data set.}
 \label{fig:MD_trainValid}
\end{figure}


\begin{figure}[!t]\centering
\includegraphics[width=.7\linewidth]{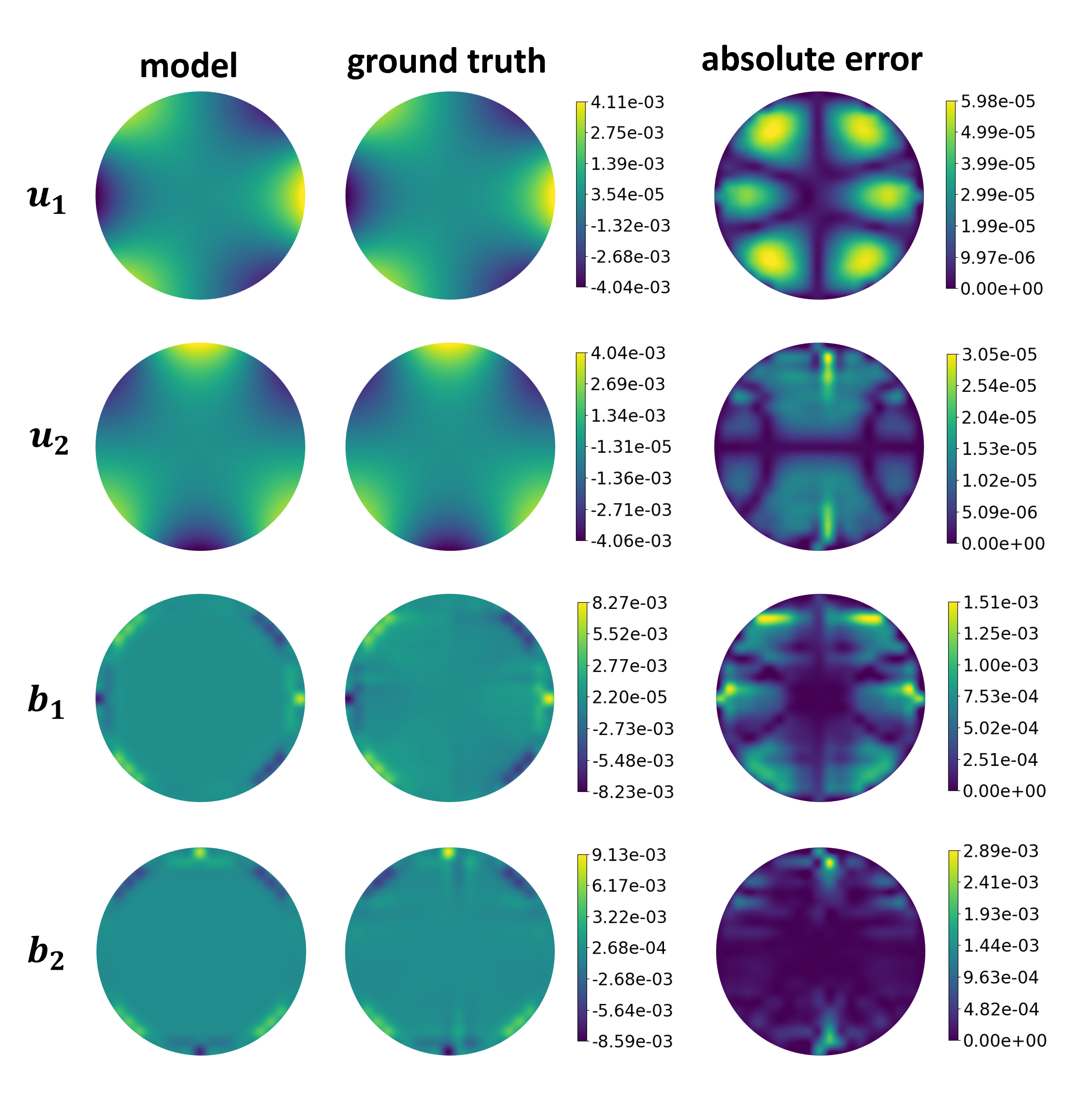}
 \caption{PNO prediction results of Example I: Graphene. Comparison of the displacement field (given body force and boundary conditions) and the prediction of body force (given displacement field) versus the corresponding ground truth for a test sample.}
 \label{fig:MD_test}
\end{figure}

In this example we use PNO to learn a continuum constitutive model for single layer graphene at zero temperature, using synthetic data sets generated by molecular dynamics (MD) simulations.

Graphene is a two-dimensional form of carbon arranged in hexagonal lattice structure, with excellent mechanical properties and with a variety of applications \cite{akinwande2017review}. 
The MD simulation uses the Tersoff interatomic potential \cite{Tersoff1988} to compute the forces between the atoms.
The MD code finds the equilibrium
displacements of atoms subjected to predefined external force fields and boundary conditions. 
The details of the MD data generation can be found in \cite{silling2023peridynamic, you2022data}. 
Here, a summary of these data sets is provided with a focus on features relevant to the present study.
The training and validation sets, with 70 and 10 samples respectively, model a 100\AA\,$\times$\,100\AA\,2D square region, consisting of hexagonally arranged atoms with an interatomic distance of 1.46\AA\, and with periodic boundary conditions. 
The external forces in the training and validation
sets are prescribed oscillatory functions of position with different wavelengths for each sample. 
The validation set uses external forces in both spatial directions that are significantly different from those used in the training set. The test set includes 4 samples on a disk of radius 100\AA\, with a traction-free boundary. 
The external forces are applied on an exterior ring with 50\AA\, distance from the outer edge, with zero force on the disk
that is inside the ring. 
Figure \ref{fig:MD_samples} shows examples of the training, validation, and test samples using MD. 
A coarse-graining method is applied to the MD data to obtain the displacements and forces on a 
uniform rectangular grid with a nodal spacing of $\Delta x=$5\AA,
which is coarser than the original MD lattice. 
Figure \ref{fig:MD_CG} shows the hexagonal lattice of atoms, the MD domain, and the resulting 21$\times$21 coarse-grained domain for training and validation sets. 
The radius for the smoothing functions used in the coarse-graining method is chosen to be 10\AA, that is, twice the coarse grained grid spacing. 
The smoothing function used in the coarse graining algorithm determines the horizon $\delta$ in the coarse grained data.
The coarse grained displacements no longer represent individual atoms, but instead represent a weighted average of their motions.
More details of the coarse-graining method can be found in \cite{silling2023peridynamic, you2022data}. 
The displacement and external forces for typical training and validation samples are shown in Figure \ref{fig:MD_trainValid}.

The training and validation data sets in this problem had nonzero forces and followed the Case I procedure described in Algorithms \ref{alg:ML}. 
After training several PNO models with a range of hyperparameters in the optimizer, 
the best model was selected to the minimize the average error \eqref{eqn: caseI_loss} in the training and validation data sets. 
After selecting the best model, it was applied to the test set, which, as noted above, is significantly different from the training set. 
Since graphene is isotropic at small deformations, the kernel function (\ref{eqn:omega_iso}) was used. 
The MLP for $\omega^{NN}$ in the trained model had widths (1, 256, 512, 1), while the $\sigma^{NN}$ MLP widths were (4, 512, 512, 1). 
The horizon size is taken as $\delta=4\Delta x$=20\AA, as suggested in \cite{you2022data}. 
The errors in both forces and displacement were computed for all three data sets.

As noted in that algorithm, two different errors were reported to show the accuracy of the trained PNO: 
1) the error in the predicted internal force for a given the displacement field; and 
2) the error in the predicted displacement for a given combination of the boundary conditions and external forces. 
For training and validation sets, the Dirichlet BC in $\omg_I$ were obtained by extending the prescribed displacement field from the original 100\AA\,$\times$100\AA\, domain into the ring ($\omg_I$) using the periodic boundary condition. 
For the test set, since the domain is circular and not periodic, the solver computed the displacements on the interior disk with the radius of 50\AA, treating the displacement on the exterior ring with the thickness of $2\delta$ as the Dirichlet BC.

\begin{figure*}[h!]
  \begin{minipage}[b]{0.72\linewidth}
    \centering%
   \begin{tabular}{|c|c|c|c|c|}
\hline
\multicolumn{2}{|c|}{error} & training (square) & validation (square) & test(disk) \\
\hline
\multirow{2}{*}{force} & LPS & 9.81\% & 13.28\% & 2.03e-1 \\
 & PNO & {7.89}\% & {10.23}\% &{4.97e-3} \\
\hline
\multirow{2}{*}{displacement} & LPS & 11.72\% & 7.16\% & 6.75\% \\
 & PNO & {10.00}\% & {2.78}\% & {3.80}\%  \\
\hline
\end{tabular}
\par\vspace{20pt}
      \end{minipage}
\begin{minipage}[b]{0.27\linewidth}
    \centering
\subfigure{\includegraphics[width=1.\linewidth]{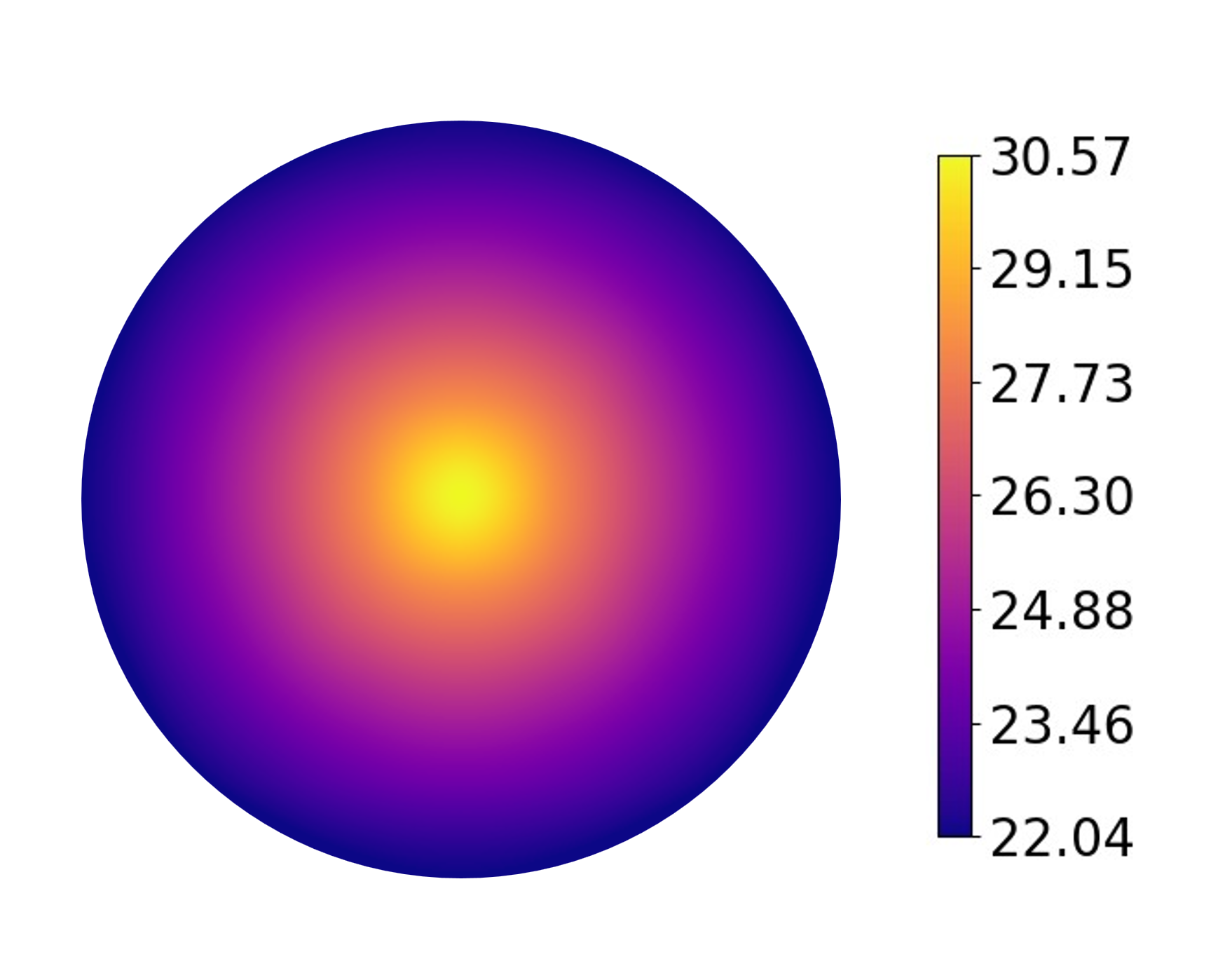}}
    \par\vspace{0pt}
  \end{minipage}%
\caption{PNO prediction results of Example I: Graphene. Left: $L^2$-errors of the proposed PNO model, in comparison with the optimal LPS model \cite{you2022data}. Right: The learned kernel function $\omega^{NN}(|\xib|/\delta)$ of the PNO model.}
\label{fig:MD_kernel}
\end{figure*}

During training, we follow the Case I procedure and use the relative $L^2$ error of forces as the training and validation criteria. In the nonlinear solver, we initialize the displacement field as $\ub^{s}_0=0$ for all cases. Then, we compute the relative $L^2$ errors for training, validation, and test sets\footnote{Since the external forces on the interior disk were zero, on the test set we report the absolute error for the forces.}. The predicted displacements and predicted forces are compared against ground-truth values for a test sample in Figure~\ref{fig:MD_test}. In the left of Figure \ref{fig:MD_kernel} we report the computed errors from the trained PNO, compared with those from an optimal LPS model, 
{\emph{i.e.,}} a peridynamic model of the form \eqref{eqn-lps} with an algebraic kernel function optimized on the same training and validation data sets \cite{you2022data}. 
One can clearly see the improvement of the present PNO model over the previous data-driven approach where the focus was only on learning a particular kernel function, $\omega$. 
The PNO allows the force state function to be learned from the data while adhering to the fundamental structure of an ordinary-state-based peridynamics and hence having material objectivity and the balance of angular momentum guaranteed. This comparison highlights the importance of learning the model form in terms of the force state $\ut$ together with the kernel function $\omega$: the test error for forces has been improved by $97.6\%$, and the error for displacements has been improved by $43.7\%$. Further more, in the right plot of 
Figure \ref{fig:MD_kernel} we show the learned kernel function for graphene. 
While for this example $\omega^{NN}$ is a 1D function (only a function of bond length), it is shown in a 2D image
to demonstrate how the values vary for neighbor points in the physical domain. 
Interestingly, this learned kernel shares some important features, 
positive values with the maximum in center and a monotonic decay in radial direction, 
with qualitative forms for the influence states that are typically assumed 
in the peridynamics literature \cite{bobaru2016handbook,seleson2011role}.

\subsection{Example II: Anisotropic Hyperelastic Material}


The previous example demonstrated the learning of a continuum constitutive law from microscale simulations. The displacements in that example were in the range of small deformation and the material was known to be isotropic in this deformation regime. 
In the present example, we use the PNO to learn from the data for a large-deformation, hyperelastic, anisotropic material. 
As in the graphene example, the data in this example is also synthetic, but is generated with a classical PDE-based constitutive law and finite element method (FEM) code rather than MD. 
The sample is a 2D sheet using the Holzapfel-Gasser-Ogden (HGO) \cite{holzapfel2000new} material model in plane stress. 
The material parameters given in \cite{you2022physics} are used in the HGO model.
The PNO for this material uses the kernel in the form of \eqref{eqn:omega_aniso} which allows anisotropy through dependency on the bond vector, containing information on magnitude and direction.
The material contains fibers that are aligned in the vertical ($x_2$) direction, making the response stiffer in that direction. 
This is unlike the radially symmetric form that is adequate for the graphene example, since graphene is isotropic.

\begin{figure}[!t]\centering
  \begin{minipage}[b]{0.3\linewidth}
\includegraphics[width=1\linewidth]{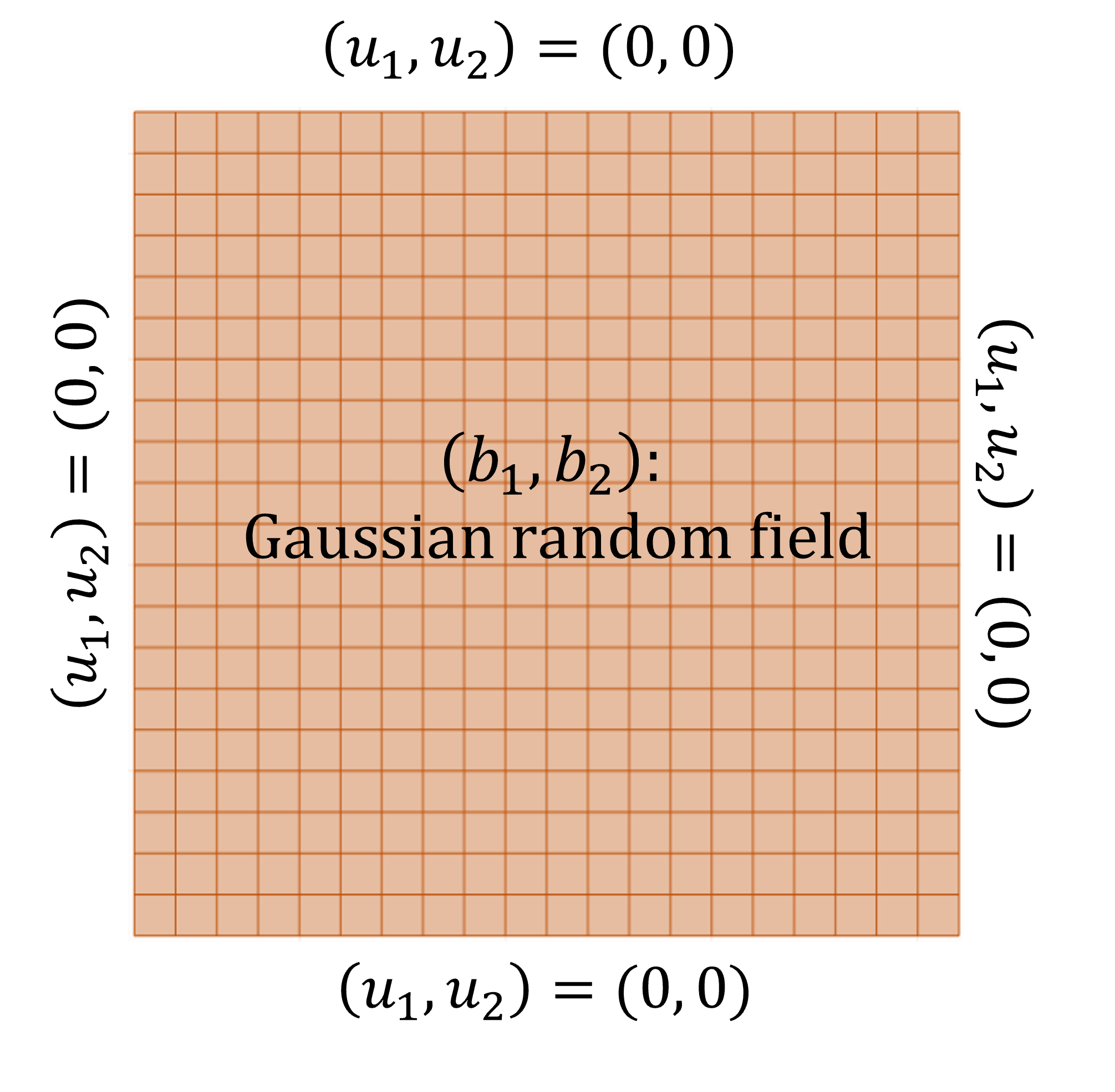}\par\vspace{30pt}
\end{minipage}
  \begin{minipage}[b]{0.6\linewidth}
\includegraphics[width=1\linewidth]{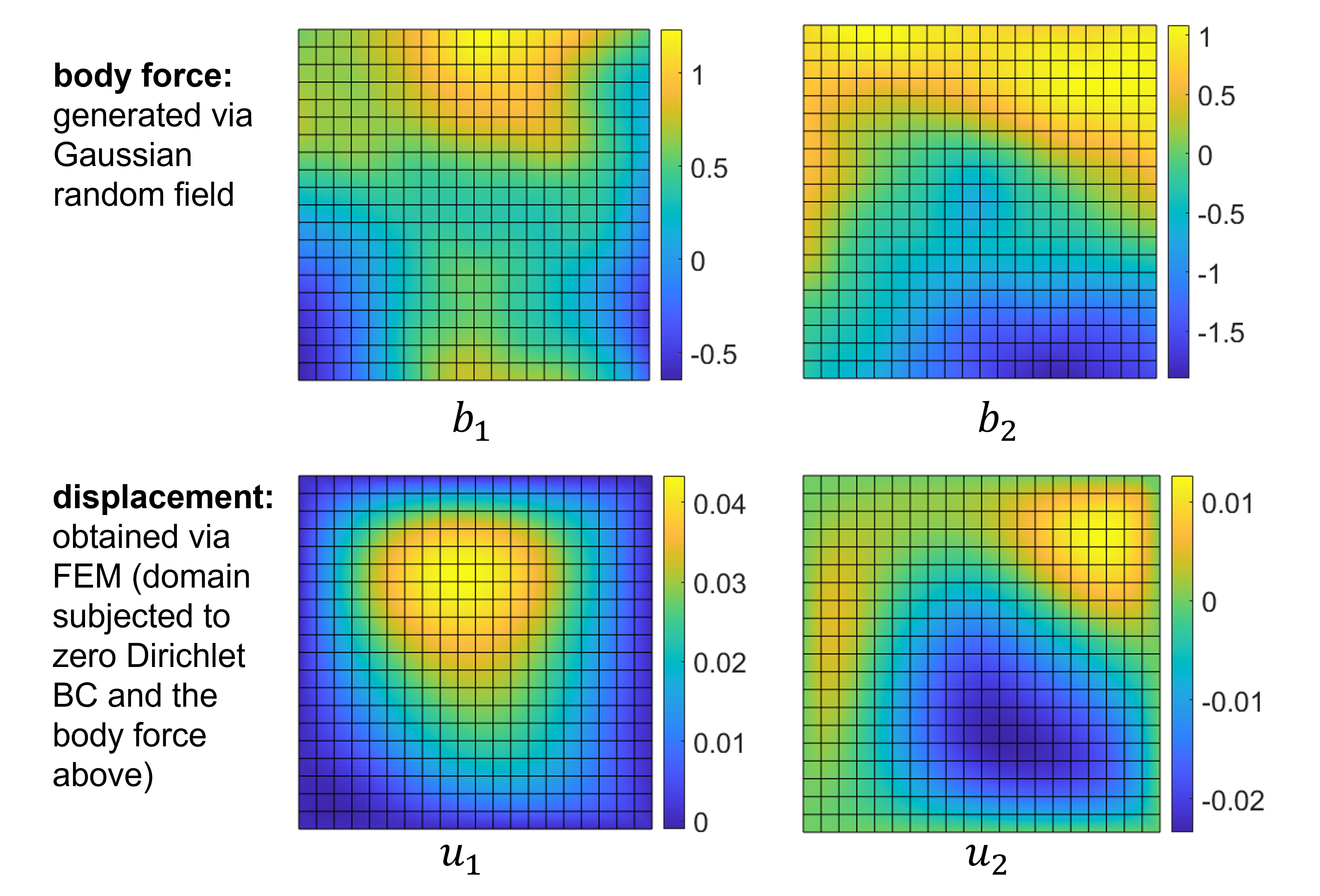}
\end{minipage}
 \caption{Settings for data generation in Example II: Anisotropic Hyperelastic Material. Left: The geometry to generate synthetic hyperelastic data using FEM and HGO material model. Right: An example of the hyperelasicty data generated via FEM and HGO material model.}
 \label{fig:HGO_setup}
\end{figure}

\begin{figure}[!t]\centering
\includegraphics[width=.5\linewidth]{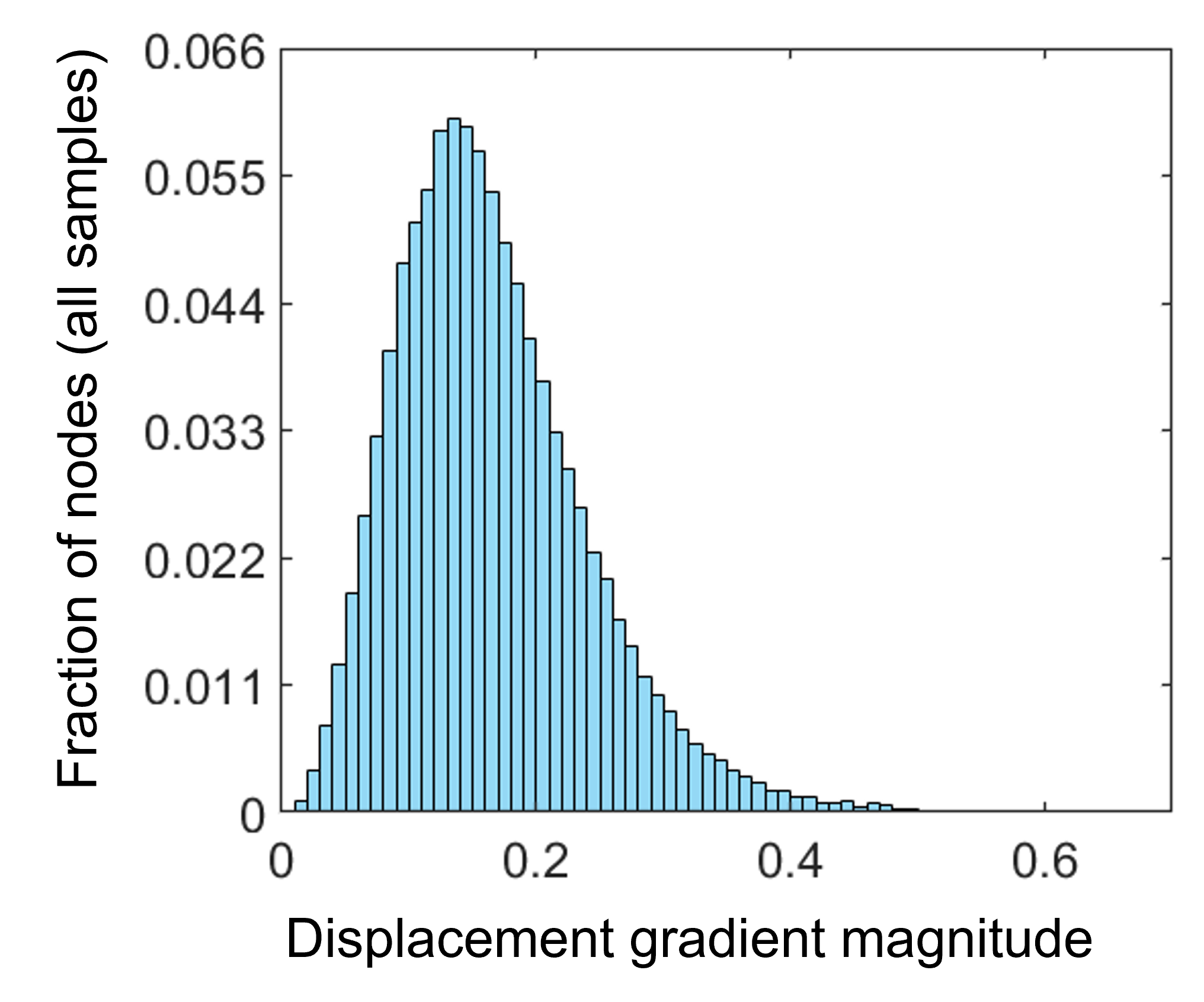}
 \caption{Distribution of in-plane displacement gradient magnitude in Example II, considering all nodes in all samples.}
 \label{fig:HGO_dispGrad}
\end{figure}

In the left plot of Figure \ref{fig:HGO_setup} we demonstrate the setting of the FEM solver used for the data generation: A $1\times 1$ square domain is uniformly meshed using 21$\times$21 nodes, and hence the grid size $\Delta x=0.05$. 
The domain is subjected to external forces generated from a Gaussian random field and zero Dirichlet BC on all the boundaries. The open-source FEM package FEniCS \cite{alnaes2015fenics} is used to solve the static problem for 250 instances. In each instance, different external loads $\bb$ are applied, and the displacement field $\ub$ is computed.
Figure \ref{fig:HGO_setup}, on the right, shows one instance of the generated data, and Figure \ref{fig:HGO_dispGrad} shows the distribution of the in-plane displacement gradient magnitude, implying a large-deformation regime.
The data is then (randomly) divided into training, validation, and testing sets with 200, 25, 25 samples for each, respectively.

\begin{figure}[!t]\centering
\includegraphics[width=.7\linewidth]{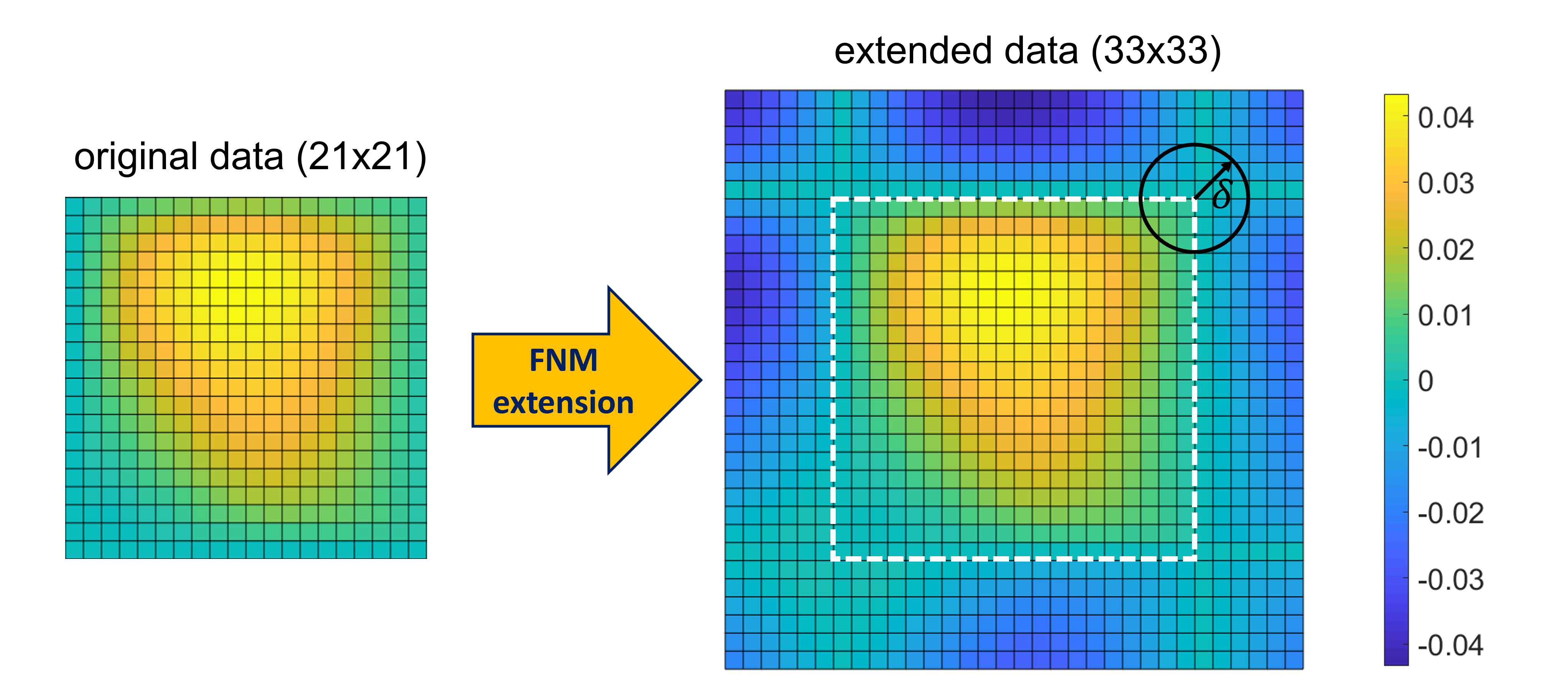}
 \caption{Demonstration for data generation in Example II: Anisotropic Hyperelastic Material. Domain extension along the boundaries to with values assigned according to the mirror-based fictitious nodes method \cite{oterkus2014peridynamic,zhao2020algorithm} to approximate local Dirichlet BC of zero.}
 \label{fig:HGO_FNM}
\end{figure}

 \begin{figure}[!t]\centering
 \includegraphics[width=.99\linewidth]{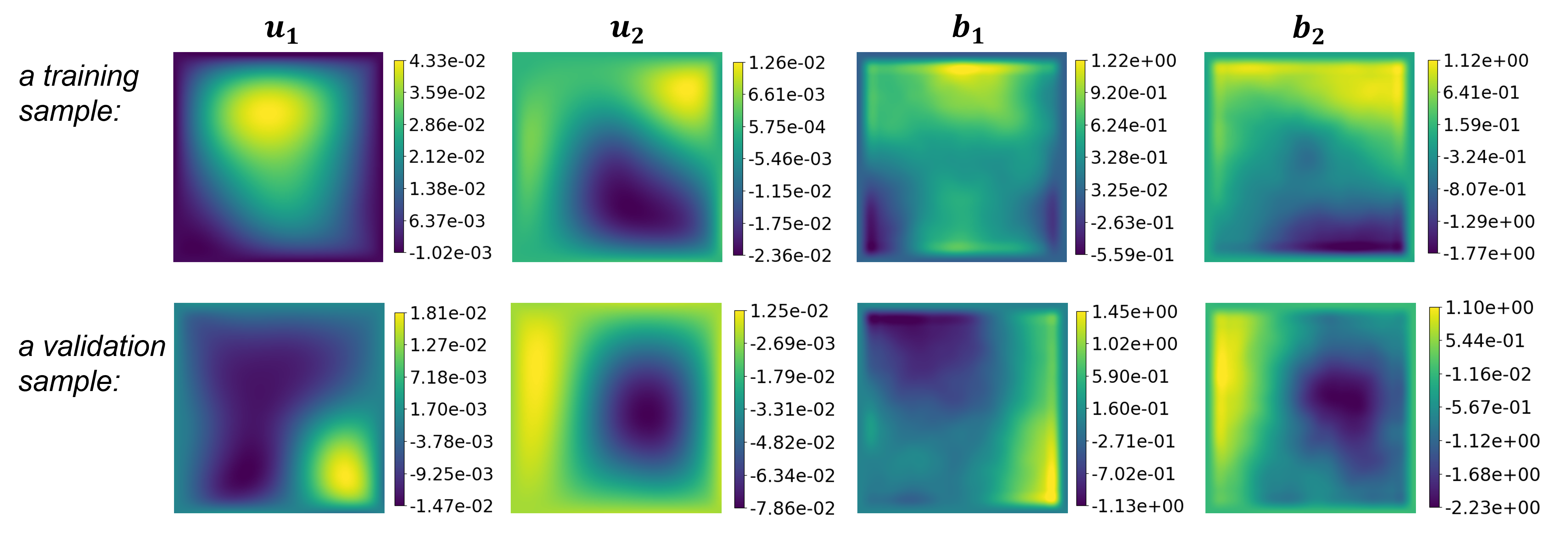}
  \caption{The external force and the corresponding displacement field components for a training and a validation sample of Example II: Anisotropic Hyperelastic Material.}
  \label{fig:HGO_trainValid}
 \end{figure}

\begin{figure}[!t]\centering
\includegraphics[width=.7\linewidth]{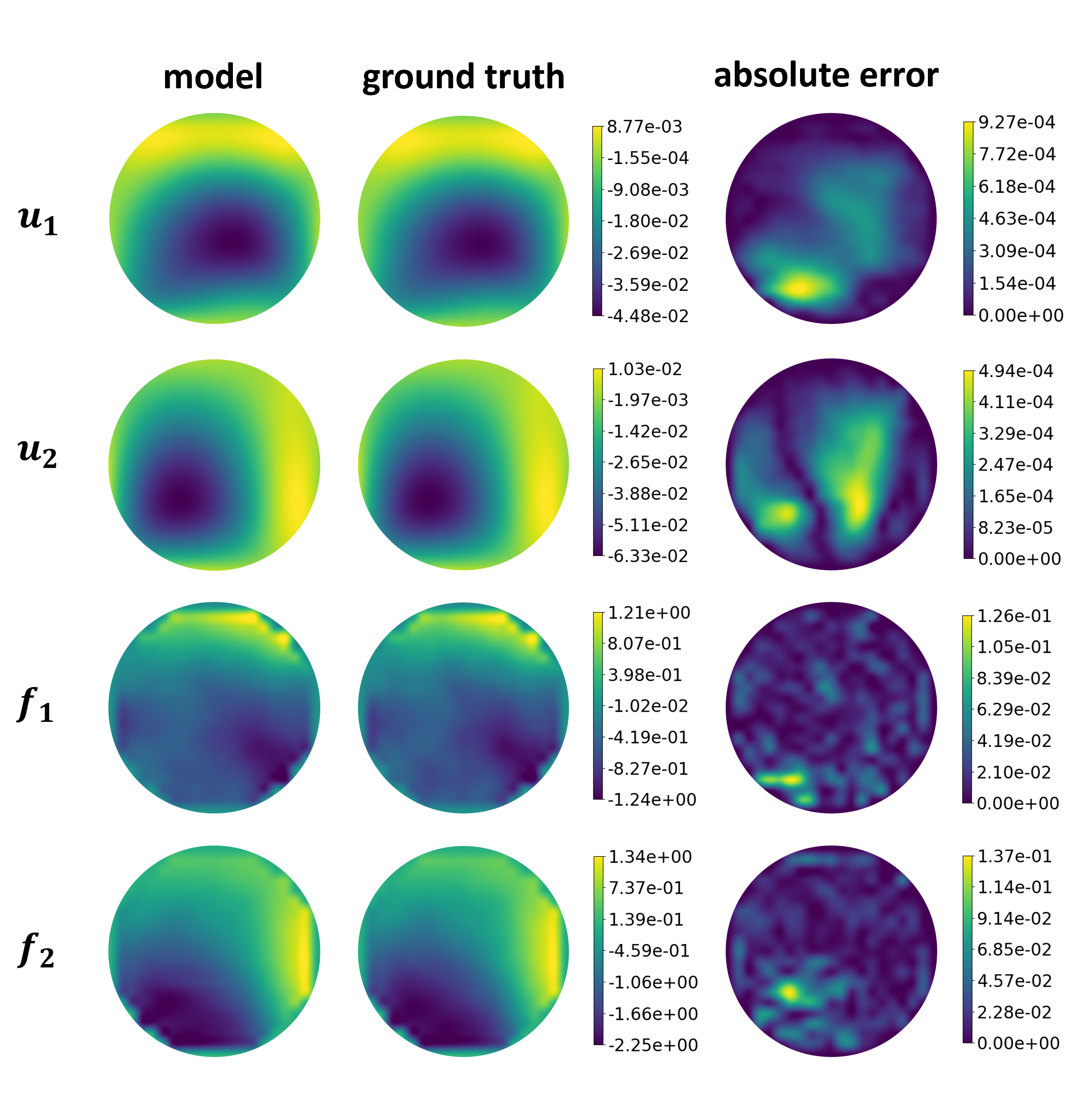}
 \caption{Demonstration of test samples in Example II: Anisotropic Hyperelastic Material. PNO prediction of the displacement field (given external force and boundary conditions) and the prediction of body force (given displacement field) versus the corresponding ground truth for a test sample.}
 \label{fig:HGO_test}
\end{figure}


In the present example, we consider $1\times 1$ square domains for the training and validation samples (as demonstrated in Figure \ref{fig:HGO_trainValid}), all with (local) Dirichlet boundary conditions. To supply the required nonlocal Dirichlet data in the boundary volume $\omg_I$, a mirror-based fictitious nodes method (FNM) \cite{oterkus2014peridynamic,zhao2020algorithm,foss2023convergence} is used to assign values within the extended boundary. 
The mirror-based FNM provides volume constraints on the extensions such that the desired local BC is well approximated on the original local boundary. 
To do this, the following formula is used:
\begin{equation}
\ub(\xb_{vc}) = 2\ub(\xb_{bc}) - \ub(\xb_{mir})\text{ ,}
\end{equation}
where $\xb_{vc}$ is a constrained node in the extension. 
$\xb_{bc}$ denotes a point on the local boundary that is closest to $\xb_{vc}$, and $\xb_{mir}:=2\xb_{bc}-\xb_{vc}$ is the ``mirror node'' for $\xb_{vc}$ inside the domain across the boundary. 
$\xb_{vc}, \xb_{bc},$ and $\xb_{mir}$ form a straight line normal to the boundary with $\xb_{bc}$ in the middle. 
More details of this method can be found in \cite{zhao2020algorithm}. 
Figure \ref{fig:HGO_FNM} shows the nonlocal BC in a $2\delta$-thick extended layer for the $u_1$ component of the sample shown in Figure \ref{fig:HGO_setup}. 
While using an extended region to supply nonlocal boundary data, 
the PNO method for purposes of evaluating errors is restricted to the interior.

To demonstrate the generalizability of PNOs to different geometries and BCs, we used a disk-shaped region rather than a square for the test set. 
The central disk had a radius of $0.5$ (Figure \ref{fig:HGO_test}).
A ring with thickness $2\delta$-thick was used as an extension of the region for the nonlocal BC, as in the graphene example.
Within the ring, displacements from FEM were presecribed, and the PNO solution was then obtained within the disk. The widths of the MLPs for $\omega^{NN}$ and $\sigma^{NN}$ are respectively: (2, 256, 512, 1) and (4, 512, 512, 1). 
Since the external forces are nonzero, the training and testing follow the algorithm and formulations of Case I described in Section \ref{sec:alg}.

\begin{figure*}[h!]
  \begin{minipage}[b]{0.72\linewidth}
    \centering%
\begin{tabular}{|c|c|c|c|c|}
\hline
 PD horizon size &error& training & validation & test \\
\hline
\multirow{2}{*}{$\delta = 3\Delta{x}$}&force & 3.21\% & 3.68\% & 3.51\% \\
&displacement & 0.73\% & 0.77\% & 0.72\% \\
\hline
\multirow{2}{*}{$\delta = 4\Delta{x}$}&force & 3.93\% & 4.45\% & 4.37\% \\
&displacement & 0.90\% & 0.90\% & 1.03\% \\
\hline
\multirow{2}{*}{$\delta = 5\Delta{x}$}&force & 3.50\% & 3.89\% & 3.82\% \\
&displacement & 0.86\% & 0.97\% & 0.98\% \\
\hline
\end{tabular}
\par\vspace{10pt}
      \end{minipage}
\begin{minipage}[b]{0.27\linewidth}
    \centering
\subfigure{\includegraphics[width=1.\linewidth]{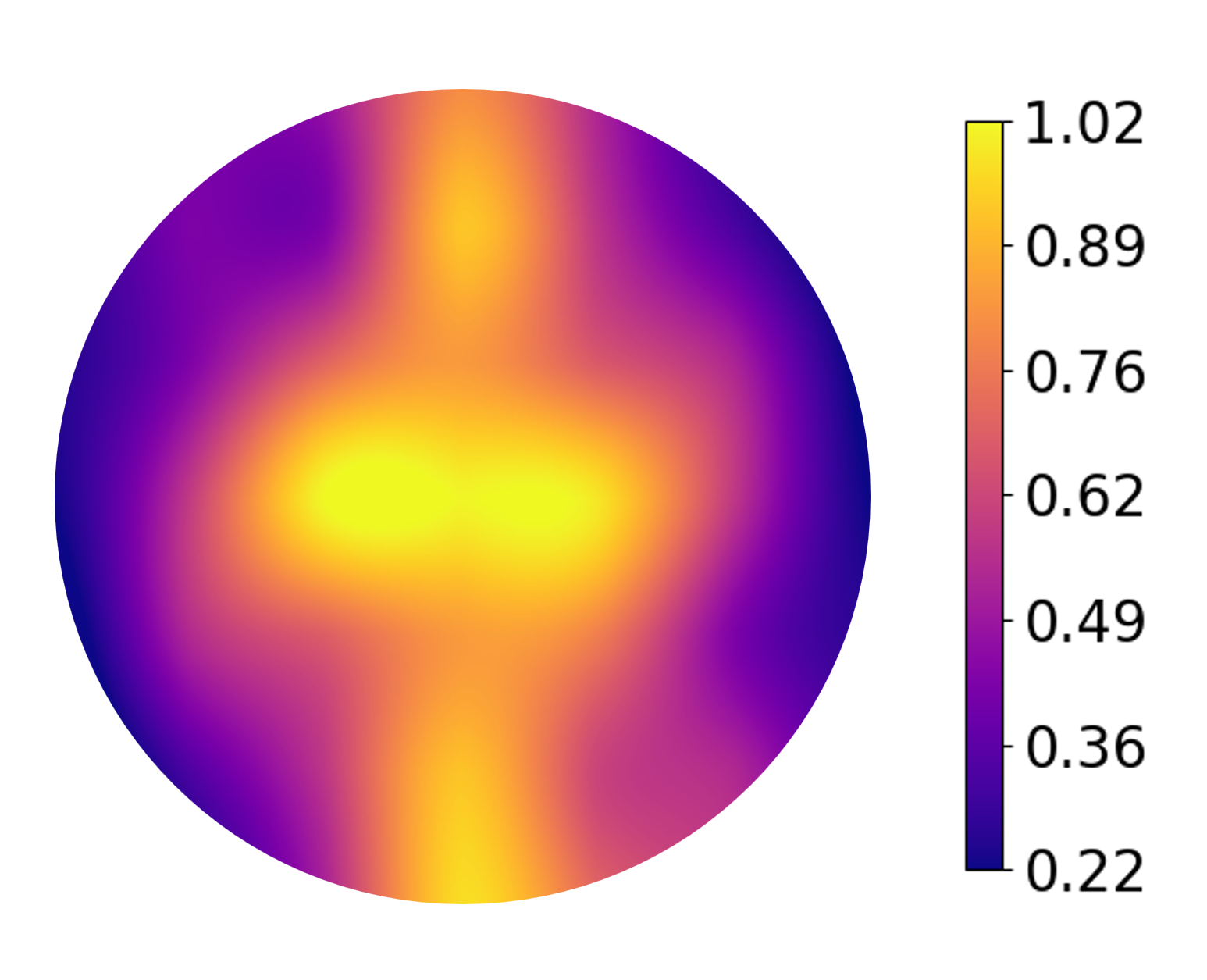}}
    \par\vspace{0pt}
  \end{minipage}%
\caption{PNO prediction results of Example II: Anisotropic Hyperelastic Material. Left: $L^2$-errors of the proposed PNO model with different horizon sizes $\delta$. Right: The learned kernel function $\omega^{NN}(\xib/\delta)$ of the PNO model.}
\label{fig:HGO_kernel}
\end{figure*}

Recall that in the graphene example, the horizon $\delta$ is determined by the coarse grained data.
In the present example, there is no coarse graining, and $\delta$ is an adjustable parameter.
To see the effect of $\delta$, the PNO is trained assuming three different horizon sizes: 
$\delta = 3\Delta x$, $4\Delta x$, and $5\Delta x$, where $\Delta x$ is the FEM mesh spacing. 
For each $\delta$, the PNO is trained on the same training set, and the best model is selected based on the minimum validation error. In the left of Figure \ref{fig:HGO_kernel} we report the relative $L^2$-error for forces and displacements. Comparing the results for different $\delta$ suggests that the learning results have been compatible when using different horizon sizes, that is, there is no preferred internal length scale for the material. 
This observation is consistent with the fact that this data is generated using a local theory, that means the ground-truth solution has no length scale.
Hence, in the rest of tests we will use $\delta = 3\Delta x$, which has the lowest computational cost. In Figure \ref{fig:HGO_test}  we show the ground-truth, PNO predictions, and absolute error fields for displacements and forces for one of the test samples. One can see that the model has successfully capture important solution features visually. This fact is also verified when comparing the test error with training and validation errors: as shown in the left of Figure \ref{fig:HGO_kernel}, all three errors are in a similar scale, highlighting the generalizability of PNOs. To further investigate the physical meaning of the learning results, the learned kernel function, $\omega$, for this example is plotted in the right plot of Figure \ref{fig:HGO_kernel}. 
The kernel values are generally larger in vertical direction, reflecting the stiffer response in the $x_2$-direction
due to the anisotropy in the HGO model.

\subsection{Example III: A Biological Tissue}
\label{subsec-example3}

\begin{figure}[!t]\centering
\includegraphics[width=.99\linewidth]{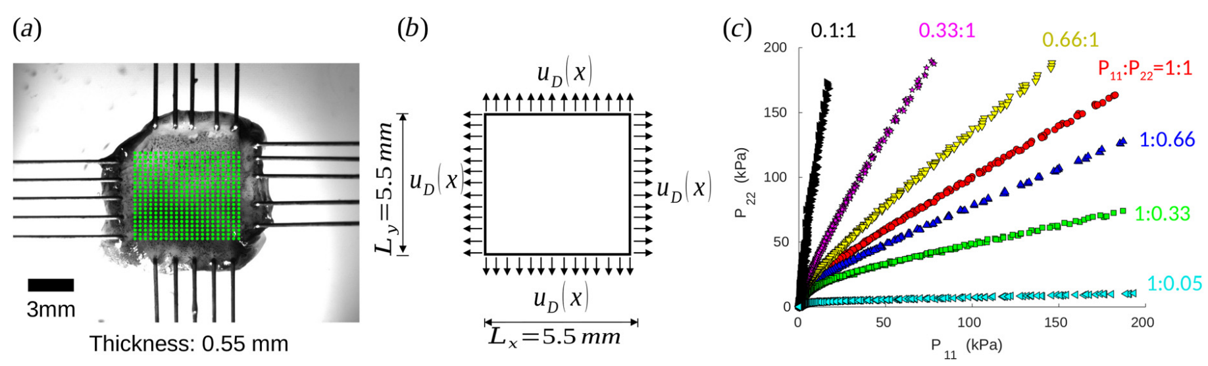}
 \caption{The experimental set up for biaxial testing of the biological tissue \cite{you2022physics} in Example III.}
 \label{fig:tissue_setup}
\end{figure}

In the present example, the PNO is applied to learn a constitutive model for a real material based on data collected from experimental measurements \cite{you2022physics}. In particular, we consider a porcine TVAL specimen subjected to biaxial stretching with digital image correlation (DIC) for recording the displacement field during the tests. 
The material is heterogeneous and anisotropic and, in these tests, undergoes strains as large as $90\%$. 
Figure \ref{fig:tissue_setup} shows the experimental setup (a), the corresponding boundary-value problem (b), and the biaxial loading ratios (c). 
The tissue undergoes seven different protocols (biaxial loading ratios). 
The original experiments repeated each protocol for three cycles. 
However, in the present study we aim to compare the PNO against the best available continuum models, 
and the results of such models are available only for the first cycle of each protocol \cite{you2022physics}. 
The training set used for this problem consists of 100 samples equi-spaced temporally over the total measurement sequence of 7378 samples from all seven protocols. 
The validation set consists of 20 samples from the remainder of the data, such that samples with high, low, and mid-level stretches from all seven protocols are included. 
For testing, and to compare the accuracy of the PNO against different classical (local) constitutive models, the same 1455 test samples are employed that were used in \cite{you2022physics}

The displacements are measured in each sample with DIC at 441 positions on a 21$\times$21 grid in a square $0.55mm\times 0.55mm$ domain with uniform spacing $\Delta x=0.0275mm$.
The experiments are quasi-static with zero external forces. 
Therefore, Case II (see Section~\ref{sec:alg}) is used for training and testing.
Recall that in Case II, a nonlinear solver is required in each sample and each epoch, which can be computationally expensive.
To reduce the computational cost, values of {\emph{tol}}=0.001 and {\emph{itr}}$_{max}$=15 in Algorithm~\ref{alg:ML} are used with the training set,
which are relatively relaxed criteria.
To make 15 allowed iterations sufficient to converge for most samples, the initial displacements are obtained from 
other samples that are close in the measurement sequence. 
Also, smaller neural networks are used for the PNO than in the previous two examples. 
The MLP widths are set to (2, 32, 64, 1) for  $\omega^{NN}$, and (4, 64, 64, 1) for $\sigma^{NN}$. 
To allow for anisotropy, (\ref{eqn:omega_aniso}) is used as the form of $\omega^{NN}$. For the validation set, since they do not go under auto-differentiation, the GPU memory is not affected by the solver's iterations. Hence, we used zero initial displacement and \emph{itr}$_{max}$=100 for the validation set to filter out models that are overfitted to close initialization during training.

Although the tissue is slightly heterogeneous, the PNO in this study is homogeneous,
that is, no part of the operator is explicitly a function of $\xb$.
Since the length scale of the heterogeneities in the samples is not known, the largest horizon size that is practical to use
in GPU memory was assumed, which is $\delta = 5\Delta x$.
As in the previous example, we use the mirror-based FNM \cite{oterkus2014peridynamic, zhao2020algorithm} to supply boundary conditions on a $2\delta$-extension on the original $0.55mm\times 0.55mm$ domain.  
\begin{table*}[ht!]
\caption{Learning results of Example III: A Biological Tissue: average relative error of the displacement field for the biological tissue under seven protocols of biaxial tension test.}
\label{tab:tissue_train}
\vskip 0.15in
\begin{center}
{\centering
\begin{tabular}{|c|c|c|}
\hline
 training (100 samples) & validation (20 samples) & test (1455 samples) \\
\hline
 7.02\% & 9.47\% & 7.01\% \\
\hline
\end{tabular}}
\end{center}
\vskip -0.1in
\end{table*}

\begin{figure}[!t]\centering
  \begin{minipage}[b]{0.65\linewidth}
\includegraphics[width=1\linewidth]{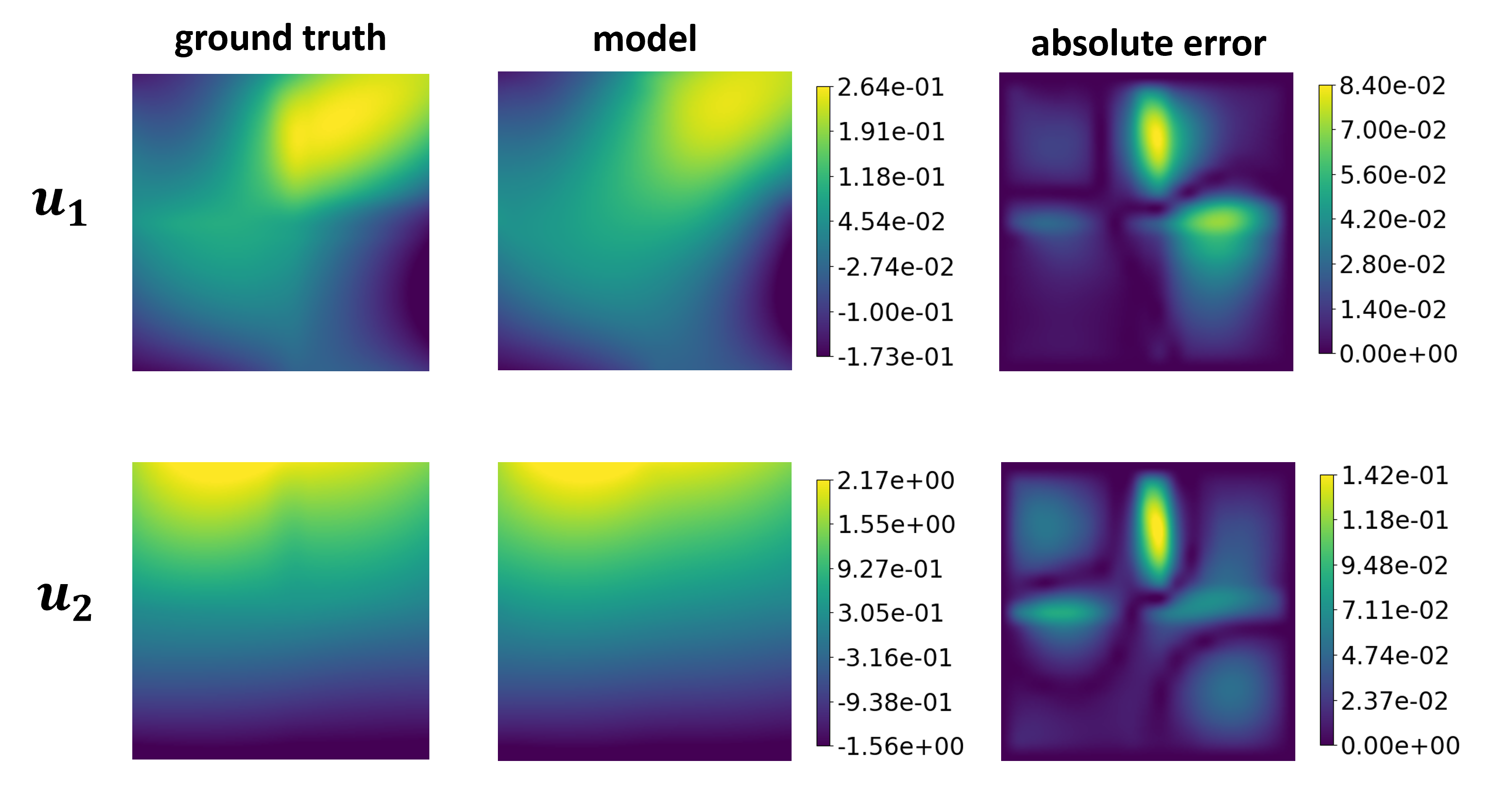}
\end{minipage}
  \begin{minipage}[b]{0.33\linewidth}
\includegraphics[width=1\linewidth]{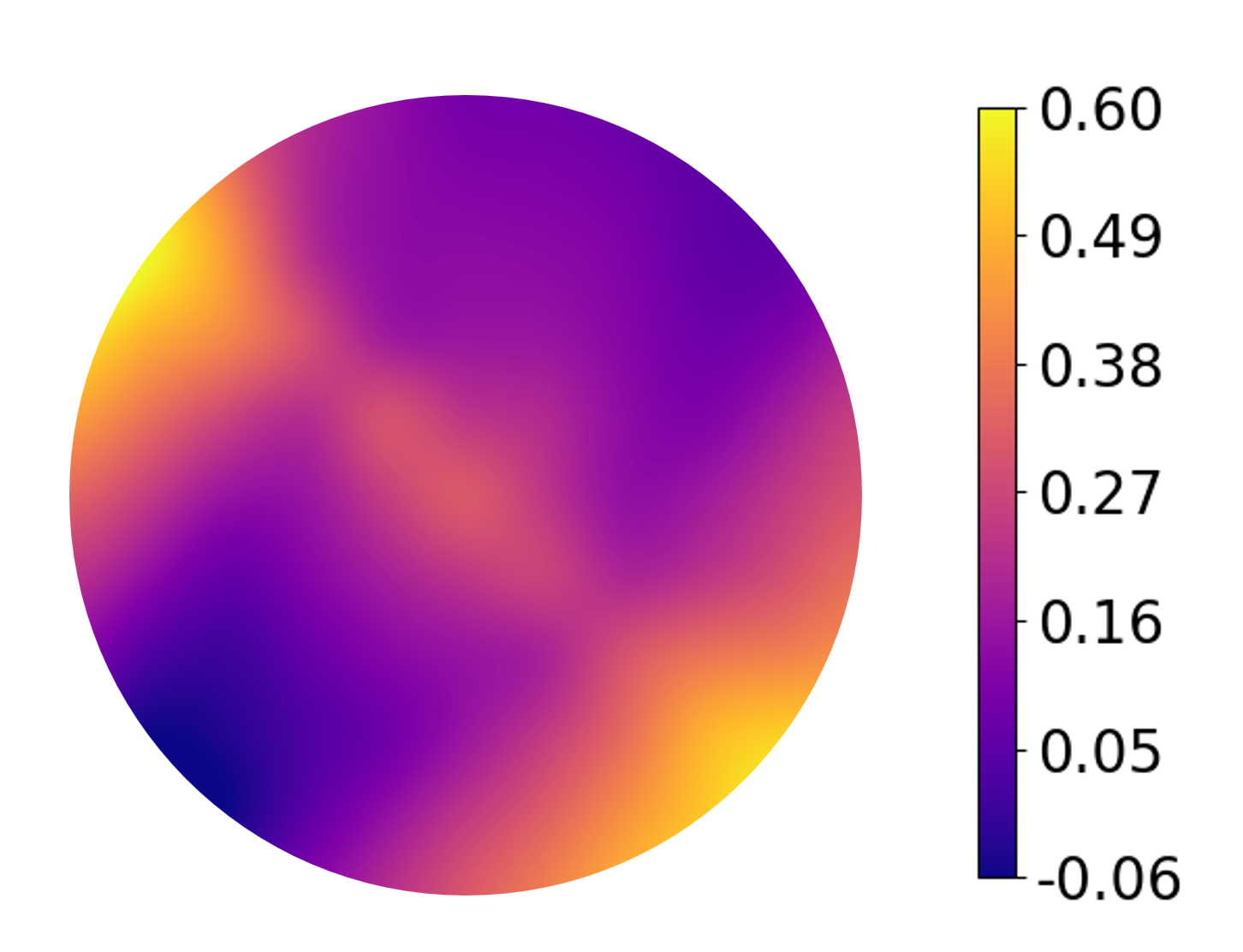}\par\vspace{30pt}
\end{minipage}
 \caption{Learning results of Example III: A Biological Tissue. Left: PNO predictions of the displacement field versus ground truth (experimentally measured) for a test sample. Right: The kernel function $\omega^{NN}(\xib/\delta)$ learned in the PNO model.}
 \label{fig:tissue_results}
\end{figure}

Table \ref{tab:tissue_train} shows the average relative $L^2$ displacement errors for the training, validation and test sets. Then, on the left plot of Figure \ref{fig:tissue_results} we show one instance of the test set. The ground-truth displacement field, the PNO prediction, and the absolute errors are plotted. 
The PNO model reproduces the material response very well except at specific locations.
A possible explanation for this is that heterogeity in the sample is strongest at these locations, so the homogeneous model is less accurate there.
Another is that in the experiment, there is some unavoidable nonuniformity in the way the loads are applied.
This can be seen in the left image in Figure~\ref{fig:tissue_setup}, which shows some variation in the spacing of the
rod-like structures that load the edges of the sample. To further investigate the physical meaning of our learning results, on the right plot of Figure \ref{fig:tissue_results} the learned kernel is provided. The image shows that this kernel represents an anisotropic material with stiffer responses in -45/+135$^o$ direction. 
This type of tissue is reported to have two different families of collagen fibers in its microstructure \cite{fitzpatrick2022ex}.
Along with possible heterogeneity, this 
may explain the apparently more complex dependence of the kernel on the bond vector, compared with the previous example.


For comparisons with the PNO approach, we employed three models for the planar stress--strain behavior of tissues: (i) a Fung-type model \cite{fung1993remodeling}, (ii) an invariant-based model \cite{johnson2021parameterization,lee2014inverse,kamensky2018contact}, and (iii) a structure-informed model \cite{fan2014simulation,he2021manifold,lee2015effects,lee2017vivo}. In all three models we solved for the displacement field using Abaqus \cite{abaqus2011abaqus}, and calculated their relative error with respect to the experimentally-retrieved displacements of each node. The optimal model parameters were obtained by minimizing the total relative displacement error on all training samples. That means, the training procedure is more favorable to these three baseline models, since their parameters are optimized based on all 7378 training samples while our PNO has only considered 120 training and validation samples. Further details of these models are given in \cite{you2022physics}. 

\begin{figure*}[h!]
\begin{minipage}[b]{0.65\linewidth}
    \centering
\subfigure{\includegraphics[width=1.\linewidth]{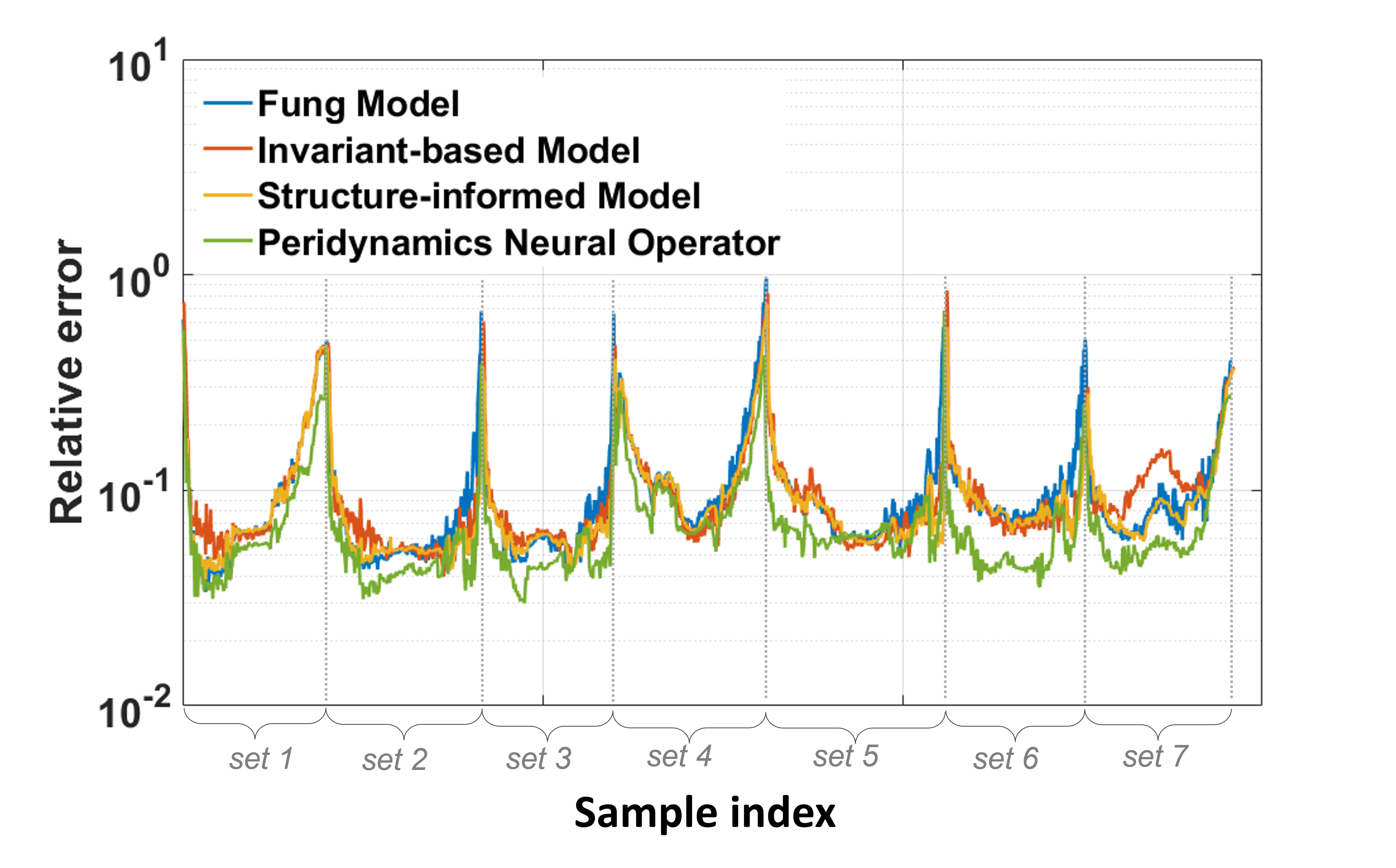}}
    \par\vspace{0pt}
  \end{minipage}%
  \begin{minipage}[b]{0.3\linewidth}
    \centering%
\begin{tabular}{|c|c|}
 \hline
 \textbf{material model} & \textbf{test error}\\
 \hline
 Fung & 10.83\% \\
 \hline
 Invariant-based & 10.73\% \\
 \hline
 Structure-informed & 9.93\% \\
 \hline
 PNO & \textbf{7.01\%} \\
\hline    
\end{tabular}
\par\vspace{100pt}
      \end{minipage}
\caption{Learning results of Example III: A Biological Tissue, and comparison with state-of-the-art classical tissue models. Left: Relative error plotted vs sample index for 1455 samples as the test set for three classical models and the learned PNO model. Right: Comparison of the average $L^2$-relative error of the displacement field over the 1455 test samples.}
\label{fig:tissue_compare}
\end{figure*}

Figure \ref{fig:tissue_compare}, on the left, shows the per-sample relative $L^2$-errors plotted for the PNO and these three classical constitutive models. The corresponding average relative $L^2$-errors are provided in  the right of Figure \ref{fig:tissue_compare}. Both results show that PNO outperforms all three of the classical models. To explain the improvement, we note that while the classical models are sophisticated and specifically designed for tissue materials, they suffer from intentional and unintentional simplifications that ignore subtle characteristics hidden in the real material. In contrast, the PNO uses the power of neural networks to learn the material model from experimental data, unconstrained by an assumed algebraic form of the material model.
Another advantage of the PNO comes from its nonlocality, which captures some aspects of the $\delta$-scale heterogeneity of the tissue.  This example has demonstrated that the PNO can learn a constitutive model for a real, heterogeneous, anisotropic, large-deformation material and that this model can outperform state-of-the-art classical constitutive models.

\section{Summary and Future Directions}\label{sec:conclusion}
In this paper, we introduced a novel nonlocal neural operator architecture for material constitutive modeling that can be trained directly from full-field spatial measurements. 
The operator is implemented within the framework of an ordinary state-based peridynamic material model.
The heart of the architecture is a neural operator with two shallow neural networks: 
the nonlocal kernel $\omega^{NN}$ together with a nonlocal model form $\sigma^{NN}$ for the peridynamic scalar force state.

The model determines the peridynamic force state vector function for any deformation. 
Because it is implemented in this way, the model automatically satisfies the conservation laws and required invariances that
peridynamics does.
It also provides the expressivity inherent in neural networks. 

As discussed in the previous sections, extensive numerical experiments were performed,
covering a variety of material modeling problems.
These include the learning of a material model from molecular dynamics simulations, 
PDE-based synthetic data sets, 
and digital image correlation (DIC) measurements. 
Empirical results show that the PNO provides physically consistent predictions for 
highly nonlinear and anisotropic materials, even in small and noisy data regimes. 
Moreover, it was verified that the learned model is generalizable to different loading scenarios and domain settings.

In future work we plan to explore how additional fundamental aspects of continuum mechanics, 
such as the polyconvexity within the context of hyperelasticity can be incorported into the PNO framework. 
We also plan to explore how the expressivity in neural operators can be exploited to better model heterogeneous media. 
It seems possible to embed the observable or trainable material heterogeneity into the model. 
Since peridynamics has been widely adopted in the engineering community because of its capability in modeling material fracture, as another natural extension we plan to extend our algorithm to material damage problems.


\section*{Declaration of Competing Interest}

The authors declare that they have no known competing financial interests or personal relationships that could have appeared to influence the work reported in this paper.

\section*{Acknowledgements}

S. Jafarzadeh would like to acknowledge support by the AFOSR grant FA9550-22-1-0197, and Y. Yu would like to acknowledge support by the National Science Foundation under award DMS-1753031. Portions of this research were conducted on Lehigh University's Research Computing infrastructure partially supported by NSF Award 2019035. 
 
This article has been authored by an employee of National Technology and Engineering Solutions of Sandia, LLC under Contract No. DE-NA0003525 with the U.S. Department of Energy (DOE). The employee owns all right, title and interest in and to the article and is solely responsible for its contents. The United States Government retains and the publisher, by accepting the article for publication, acknowledges that the United States Government retains a non-exclusive, paid-up, irrevocable, world-wide license to publish or reproduce the published form of this article or allow others to do so, for United States Government purposes. The DOE will provide public access to these results of federally sponsored research in accordance with the DOE Public Access Plan https://www.energy.gov/downloads/doe-public-access-plan.

\end{document}